\documentclass{iopjournal}
\usepackage{lmodern}

\usepackage{soul}
\usepackage{amsthm}
\usepackage[numbers,sort&compress]{natbib}
\usepackage{array}
\usepackage{makecell}
\usepackage{xcolor}
\usepackage{helvet}     
\usepackage{hyperref}
\usepackage{standalone}
\definecolor{linkblue}{HTML}{0B4FA6}
\newcolumntype{C}[1]{>{\centering\arraybackslash}p{#1}}
\newcolumntype{L}[1]{>{\raggedright\arraybackslash}p{#1}}

\usepackage[most]{tcolorbox}
\newtcolorbox{sectionsummary}[1]{
  colback=gray!5,
  colframe=black!50,
  boxrule=0.5pt,
  arc=2pt,
  left=6pt,
  right=6pt,
  top=6pt,
  bottom=6pt,
  fonttitle=\bfseries,
  coltitle=black,  
  title=#1
}

\theoremstyle{definition}



\usepackage{amsmath,amsthm,amsfonts,amssymb}
\usepackage{booktabs}
\usepackage{ragged2e}
\usepackage{graphicx}
\usepackage{fancyhdr}
\usepackage{float}
\usepackage{anysize}
\usepackage{titlesec}
\usepackage{enumerate}
\usepackage{xcolor}
\usepackage{bbold}
\usepackage{braket}

\usepackage{siunitx}
\usepackage{multirow}
\usepackage{array}
\usepackage{tabularx}

\definecolor{lime}{HTML}{A6CE39}
\DeclareRobustCommand{\orcidicon}{%
	\begin{tikzpicture}
	\draw[lime, fill=lime] (0,0) 
	circle [radius=0.16] 
	node[white] {{\fontfamily{qag}\selectfont \tiny ID}};
	\draw[white, fill=white] (-0.0625,0.095) 
	circle [radius=0.007];
	\end{tikzpicture}
	\hspace{-2mm}
}

\foreach \x in {A, ..., Z}{%
	\expandafter\xdef\csname orcid\x\endcsname{\noexpand\href{https://orcid.org/\csname orcidauthor\x\endcsname}{\noexpand\orcidicon}}
}

\usepackage[normalem]{ulem}

\newcommand{\Nst}{N^{*}}
\begin{document}
\justifying

\title{Cross-Platform Benchmarking of GHZ-State Generation in Quantum Processors via Cloud Services}

\author{Sergio Hern\'andez$^{1}$\orcidH, Unai Aseguinolaza$^{2}$\orcidF{}, David Aguirre$^{3,4}$\orcidB{}, Marta Gili$^{5}$\orcidM{}, Joseba Hern\'andez$^{5}$\orcidJ{}, Guillermo Valverde$^{6,9}$\orcidE{}, Uxue Ballarin Manso$^{6}$, Uxue Auzmendi$^{2}$, Iker Pastor$^{1}$\orcidK, Pablo Garc\'ia Bringas$^{1}$\orcidI, Ane Bl\'azquez Garc\'ia$^{5}$\orcidG{}, Rub\'en Pe\~na$^{4,7,*}$\orcidC{} and Mikel Sanz$^{3,4,8,*}$\orcidD{}}

\affil{$^1$University of Deusto, Avda. de las Universidades, 24, Bilbao 48007, Spain}

\affil{$^2$Applied Sciences Department, Faculty of Engineering, Mondragon Unibertsitatea, 20500 Arrasate, Spain}

\affil{$^3$Department of Physical Chemistry, University of the Basque Country UPV/EHU, Apartado 644, 48080 Bilbao, Spain}

\affil{$^4$BCAM - Basque Center for Applied Mathematics, Avd. Mazarredo 14, E48009 Bilbao, Spain}

\affil{$^5$Ikerlan Technology Research Centre, Basque Research and Technology Alliance (BRTA), J.M. Arizmediarrieta Pasealekua 2, 20500 Arrasate, Spain}

\affil{$^6$Vicomtech Foundation, Basque Research and Technology Alliance (BRTA), Donostia-San Sebastián 20009, Spain}

\affil{$^7$Galician Research and Development Center in Advanced Telecommunications (GRADIANT) Carretera do Vilar, 56-58 36214 Vigo, Spain}

\affil{$^8$EHU Quantum Center, University of the Basque Country UPV/EHU, Apartado 644, 48080 Bilbao, Spain}

\affil{$^9$Quantum Mads, Calle Larrauri 1, Edificio A, piso 3, puerta 28, 48160 Derio, Spain}
\\
\affil{$^*$Corresponding authors.}

\email{rpena@gradiant.org}
\email{mikel.sanz@ehu.es}


\begin{abstract}
\justifying{}
Quantum computing has rapidly evolved into a diverse ecosystem of hardware platforms that differ in their underlying technologies and operational characteristics. Consequently, benchmarking has become essential for identifying the strengths of quantum processors, comparing their performance, and enabling informed decisions when selecting the most appropriate quantum computing platform for specific applications. In this work, we benchmark nine commercially accessible quantum processors from five vendors, spanning superconducting and trapped-ion technologies, using the multipartite entanglement benchmark from the QuSquare suite. All experiments were performed through remote cloud-access services, with the default compilation of each provider and without error mitigation, reflecting the performance experienced by end users. The benchmark estimates the fidelity of Greenberger--Horne--Zeilinger (GHZ) states by direct fidelity estimation and searches for the largest GHZ state whose fidelity is compatible, within the statistical error, with the genuine multipartite entanglement (GME) threshold $F=1/2$. Our results reveal a substantial gap between the total number of physical qubits available on a processor and the size of the largest GHZ state passing the benchmark: the best superconducting processors reach $\Nst=27$ and $\Nst=24$ qubits, less than a quarter of their registers, whereas the 12-qubit trapped-ion processor passes with its full register. We further show that median calibration data are poor predictors of the benchmark outcome, and that the reported sizes depend strongly on the statistical decision rule: requiring certified GME reduces the number of devices that succeed at any tested size from nine to four. These results provide a baseline for future cross-platform benchmarking of quantum processors.

\end{abstract}

\section{Introduction}
\label{sec:introduction}

The rapid expansion of quantum computing has led to a proliferation of hardware architectures, each characterized by different capabilities and limitations. As quantum computing hardware continues to evolve, benchmarking has become essential not only for tracking technological progress and comparing the performance of quantum processors, but also for enabling researchers and organizations to make informed decisions when selecting the most appropriate quantum computing platform for their applications. Hardware specifications such as the number of qubits, coherence times, and gate fidelities provide only a partial characterization of a quantum processor and do not necessarily reflect its computational capabilities. Therefore, meaningful benchmarking requires performance metrics that capture these capabilities~\cite{acuaviva2026benchmarking,proctor2025benchmarking,hashim2025practical}. This need has led to a broad range of benchmarking approaches, including randomized benchmarking for characterizing gate performance~\cite{PhysRevA.77.012307}, quantum volume and its volumetric generalizations~\cite{cross2019validating,blumekohout2020volumetric,proctor2022measuring}, layer fidelity~\cite{mckay2023benchmarking}, a recent approach for measuring computational capability across different quantum computing architectures~\cite{proctor2026quops}, and application-oriented benchmark suites~\cite{lubinski2023application,tomesh2022supermarq,martiel2021benchmarking,mills2021application}, together with certification protocols that provide rigorous statistical guarantees~\cite{eisert2020certification,kliesch2021theory}.

Designing such benchmarks is particularly challenging because current quantum processors differ substantially in their underlying physical technologies, qubit connectivity, native gate sets, and system sizes. As a result, performance metrics that capture capabilities common to different quantum computing platforms are particularly valuable for cross-platform benchmarking. One such capability is the generation of genuine multipartite entanglement (GME) through the preparation of Greenberger--Horne--Zeilinger (GHZ) states~\cite{greenberger1989going}. Entanglement is one of the fundamental resources of quantum computing and plays a central role in many quantum algorithms and protocols~\cite{RevModPhys.81.865,jozsa2003role}, and GHZ states in particular are resources for quantum metrology~\cite{pezze2018quantum}, fault-tolerant protocols~\cite{knill2005quantum} and multiparty communication~\cite{espoukeh2014quantum}. Moreover, the preparation of a GHZ state requires the coherent operation of all participating qubits and is sensitive to gate errors, decoherence, readout errors, crosstalk and connectivity constraints. Consequently, GHZ-state preparation provides a meaningful probe of the multi-qubit capabilities of a quantum processor.

Several recent works have proposed GHZ-state preparation as a benchmark for evaluating quantum processors~\cite{euquantumflagshipskey,aguirre2025qusquare,tomesh2022supermarq}. In SupermarQ~\cite{tomesh2022supermarq}, however, the GHZ score is computed from the computational-basis output distribution only; it is therefore insensitive to the phase coherence of the state and cannot certify entanglement. Complementing these benchmarking proposals, numerous experimental studies have demonstrated the preparation of GHZ states on superconducting~\cite{mooney2021generation,wei2020verifying,song2019generation,engproc2026122005,qy9y-7ywp}, trapped-ion~\cite{monz2011fourteen,pogorelov2021compact,moses2023racetrack}, neutral-atom~\cite{Omran2019}, photonic~\cite{Zhong2018}, and silicon-based~\cite{takeda2021quantum,edlbauer2025atom} quantum platforms. For example, Mooney et al.~\cite{mooney2021generation} generated and verified a 27-qubit GHZ state on an IBM superconducting processor, while Moses et al.~\cite{moses2023racetrack} demonstrated a 32-qubit GHZ state with fidelity 0.82 on Quantinuum's trapped-ion race-track processor as part of a comprehensive hardware characterization. Hu et al.~\cite{qy9y-7ywp} generated a 17-qubit GHZ state with fidelity 0.6817(1) on a 31-qubit superconducting processor, certifying genuine multipartite entanglement using an entanglement witness. In a different hardware platform, Edlbauer et al.~\cite{edlbauer2025atom} used GHZ states of increasing size to benchmark the all-to-all connectivity of an 11-qubit silicon atom processor, demonstrating genuine multipartite entanglement among up to eight nuclear spins. Peev and Sivkov~\cite{engproc2026122005} recently evaluated GHZ-state preparation with up to 10 qubits on IBM's 127-qubit \texttt{ibm\_brisbane} processor. In another recent work, a 120-qubit GHZ state with fidelity 0.56(3) was reported on an IBM Heron processor~\cite{javadiabhari2025bigcats}; this result, however, relied on device-specific compilation, error detection with post-selection, dynamical decoupling and readout-error mitigation, i.e., on an expert-level optimization that is not part of the default workflow of a cloud user. However, these works were primarily designed as demonstrations on individual hardware platforms rather than as standardized benchmarking experiments, while existing cross-platform comparisons rely on other workloads~\cite{linke2017experimental,murali2019full,blinov2021comparison,zhu2022cross,montanez2025evaluating}. Consequently, direct comparison across quantum computing platforms remains difficult because different studies rely on different experimental methodologies and verification techniques: quantum state tomography scales exponentially with the number of qubits, whereas parity oscillations and multiple-quantum coherences require $\mathcal{O}(N)$ measurement settings, for a system of $N$ qubits, and a model fit~\cite{monz2011fourteen,wei2020verifying,mooney2021generation}, which makes rigorous finite-sample guarantees less direct.

To address these limitations, this work implements the multipartite entanglement benchmark from the QuSquare suite~\cite{aguirre2025qusquare} across nine commercially accessible quantum processors from five vendors, spanning two quantum computing technologies: superconducting qubits and trapped ions. The benchmark determines the largest GHZ state whose estimated fidelity satisfies a success criterion based on the GME fidelity threshold on each quantum processor. GHZ-state fidelity is estimated using Direct Fidelity Estimation (DFE)~\cite{flammia2011direct,dasilva2011practical}, whose sample complexity is independent of the number of qubits for stabilizer states. The estimated fidelity is then evaluated against the benchmark success criterion derived from the $1/2$ fidelity threshold associated with genuine multipartite entanglement~\cite{Guhne2010,Leibfried2005,toth2005detecting}. We make explicit the difference between a criterion that is compatible with GME and one that certifies it, and quantify how the benchmark outcome depends on this choice.
All experiments were performed through remote cloud quantum computing services, using the default compilation of each provider and without error mitigation, reflecting the performance experienced by end users. The evaluated hardware comprises quantum processors from IBM Quantum, IQM, Rigetti, AQT, and Oxford Quantum Circuits (OQC), the latter being the Qmio system operated by the Galicia Supercomputing Center (CESGA). Detailed hardware specifications are provided in Section~\ref{Hardware}.

Our main findings are the following. (i) On the large superconducting processors (107--156 qubits) the largest GHZ state passing the benchmark comprises at most 23\% of the register ($\Nst\leq 27$), and on the 20- and 54-qubit IQM processors 65\% and 37\%, respectively, whereas the 12-qubit trapped-ion processor passes with its full register, so that its result is limited by the register size rather than by fidelity. (ii) Device rankings are not predicted by median calibration data: on superconducting processors, the per-qubit GHZ error exceeds the median two-qubit gate error by about one order of magnitude, pointing to readout, idling and crosstalk errors and to qubit selection as dominant contributions. (iii) The outcome depends sensitively on the statistical decision rule: under the certification rule of QuSquare, $Y-\epsilon\geq 1/2$, only four of the nine processors certify GME at any of the tested sizes. (iv) Single-run results show statistically significant non-monotonic features, revealing run-to-run variability of qubit mapping and calibration beyond shot noise.

The remainder of this paper is organized as follows. Section~\ref{sec:background} reviews the relevant properties of GHZ states and their suitability as hardware benchmarks. Section~\ref{methodology} describes the benchmark methodology and implementation details. Section~\ref{Hardware} provides technical specifications of the quantum processors evaluated, and Section~\ref{Software} describes the software stack and execution pipeline. Section~\ref{sec:results} presents the experimental results, Section~\ref{sec:discussion} discusses them together with the limitations of the study, and Section~\ref{Conclusions} concludes with a summary of findings and future directions.


\section{Background}
\label{sec:background}
\subsection{Greenberger--Horne--Zeilinger States}

The Greenberger--Horne--Zeilinger (GHZ) state, first introduced in the context of quantum nonlocality~\cite{greenberger1989going}, is a paradigmatic example of a multipartite entangled quantum state. For a system of $N$ qubits, the GHZ state is defined as
\begin{equation}
    |\mathrm{GHZ}_N\rangle = \frac{1}{\sqrt{2}}\left(\ket{0}^{\otimes N} + \ket{1}^{\otimes N}\right),
    \label{eq:ghz_def}
\end{equation}
an equal superposition of the all-zero and all-one computational basis states.
This state exhibits \emph{genuine multipartite entanglement} (GME): it is not biseparable, i.e., it cannot be written as a convex mixture of states each of which is separable with respect to some bipartition of the $N$ qubits~\cite{Guhne2010,guhne2009entanglement}. The
experimental generation of GHZ states has been demonstrated on diverse physical
platforms, including trapped ions~\cite{monz2011fourteen,pogorelov2021compact,moses2023racetrack}, superconducting
circuits~\cite{mooney2021generation,song2019generation,javadiabhari2025bigcats}, Rydberg atoms~\cite{Omran2019}, photonic systems~\cite{Zhong2018} and silicon spin qubits~\cite{takeda2021quantum},
underscoring their central role as a benchmark resource in quantum information
science.

\subsection{Properties Relevant for Benchmarking}

Several structural properties make GHZ states particularly well-suited for characterizing quantum hardware. As a stabilizer state, $|\mathrm{GHZ}_N\rangle$ admits efficient preparation circuits composed exclusively of Clifford gates: typically a single Hadamard gate followed by a sequence of controlled-NOT (CNOT) or equivalent two-qubit entangling gates. This well-understood structure ensures that GHZ preparation circuits can be constructed and optimized for virtually any digital quantum processor, irrespective of native gate set or qubit technology. Moreover, the stabilizer structure of the GHZ state enables its fidelity to be estimated efficiently using Direct Fidelity Estimation (DFE), a scalable measurement protocol whose sample complexity is independent of the number of qubits~\cite{flammia2011direct,dasilva2011practical,aguirre2025qusquare}. The DFE procedure adopted in this work is detailed in Section~\ref{methodology}.

A key feature for certification purposes is the fidelity threshold: a fidelity of the prepared state with respect to the ideal GHZ state exceeding $1/2$ provides a sufficient condition for certifying genuine multipartite entanglement~\cite{sackett2000experimental,toth2005detecting,Guhne2010,Leibfried2005}. Since the condition is only sufficient, $F\leq 1/2$ does not imply the absence of GME; it only means that this fidelity witness does not detect it. In this work, this fidelity threshold is incorporated into the benchmark success criterion described in Sec.~\ref{methodology}.

\subsection{GHZ States as a Probe for Hardware Performance}

The collective structure of GHZ states makes them highly sensitive to hardware imperfections, which is precisely what renders them informative as benchmarks. Any $N$-qubit GHZ preparation circuit requires at least $N-1$ two-qubit entangling gates (in the absence of mid-circuit measurements and feed-forward); however, different circuit architectures achieve markedly different circuit depths. Sequential strategies---where entanglement propagates along a qubit chain via a cascade of CNOT gates---require $O(N)$ circuit depth, while tree-structured strategies that distribute entanglement in a binary-tree fashion achieve $O(\log N)$ depth, at the cost of requiring broader qubit connectivity~\cite{cruz2019efficient}. In practice, the connectivity graph of the target hardware determines which strategies are feasible and what routing overhead must be paid: on sparse topologies such as the heavy-hexagonal lattice of IBM's Heron processors, SWAP gates or longer entanglement paths are needed, increasing both circuit depth and total gate count beyond what ideal strategies would require, whereas all-to-all connected trapped-ion platforms can implement the logarithmic-depth strategy with no routing overhead, although in architectures in which entangling gates are applied sequentially the reduction in depth does not translate into a shorter execution time. With mid-circuit measurements and feed-forward, GHZ states can even be prepared in constant depth~\cite{baumer2024efficient,moses2023racetrack}; such dynamic circuits are outside the scope of the present benchmark. Regardless of strategy, every two-qubit gate contributes to error accumulation, and the global correlations of the GHZ state amplify the impact of any individual qubit error or decoherence event: unlike local observables that tolerate errors on individual qubits, a single decoherence event anywhere in the register can degrade the global GHZ fidelity; for spatially correlated dephasing, the coherence of the GHZ state can even decay quadratically with $N$~\cite{monz2011fourteen}.

As a result, the largest GHZ-state size $N^*$ satisfying the benchmark success criterion provides a single, hardware-comparable figure of merit that simultaneously encodes information about two-qubit gate fidelity, circuit depth tolerance, qubit coherence times, crosstalk, readout errors, and the quality of the qubit selection performed by the compiler. This architecture-sensitive dependence on connectivity and circuit strategy makes GHZ fidelity a discriminating benchmark across heterogeneous hardware platforms, enabling a meaningful cross-architecture comparison of coherent multi-qubit operation quality.

\section{Methodology}
\label{methodology}

In this section, we describe the methodology adopted to apply the multipartite entanglement benchmark proposed in Ref.~\cite{aguirre2025qusquare} across quantum processors based on different technologies. We first provide an overview of the benchmark protocol and then describe the implementation choices and adaptations required for its execution through cloud-based quantum computing services.

\subsection{Multipartite entanglement benchmark}

The multipartite entanglement benchmark aims to evaluate the capability of a quantum processor to generate genuine multipartite entanglement through GHZ-state preparation \cite{aguirre2025qusquare}. The benchmark determines the largest GHZ state for which the specified success criterion is satisfied. Since different quantum circuits can be used to prepare the GHZ state, the benchmark does not specify a unique circuit construction. Instead, the GHZ preparation circuit is adapted to the target quantum processor by selecting, among four candidate circuits, the circuit requiring the fewest native two-qubit gates after transpilation.

For each target GHZ-state size, the selected GHZ circuit is executed on the quantum hardware, and the fidelity of the prepared GHZ state is estimated using Direct Fidelity Estimation (DFE)~\cite{flammia2011direct,dasilva2011practical}. Let $\mathcal{S}_{\mathrm{GHZ}}$ denote the stabilizer group of the target GHZ state. The DFE protocol randomly samples $l$ non-identity stabilizers from $\mathcal{S}_{\mathrm{GHZ}}$, which are measured on the prepared state to construct the fidelity estimator, denoted by $Y$. The estimator has maximum estimation error $\epsilon$ and confidence level $1-\delta$, i.e., $\Pr\left(|Y-F|\leq\epsilon\right)\geq 1-\delta$, where $F=\bra{\mathrm{GHZ}_N}\rho\ket{\mathrm{GHZ}_N}$ is the fidelity of the prepared state $\rho$.

A GHZ-state fidelity greater than $0.5$ constitutes a sufficient condition for GME~\cite{Guhne2010,Leibfried2005}. Given the confidence interval $[Y-\epsilon,\,Y+\epsilon]$, three decision rules can be formulated: (i) the \emph{certification} rule $Y-\epsilon\geq 1/2$, which guarantees $F>1/2$ with probability at least $1-\delta$ and is the success criterion prescribed by QuSquare and by the EU Quantum Flagship KPI~\cite{aguirre2025qusquare,euquantumflagshipskey}; (ii) the \emph{point-estimate} rule $Y\geq 1/2$, which carries no statistical guarantee; and (iii) the \emph{non-exclusion} rule 
\begin{equation}
    Y+\epsilon \geq 0.5,
    \label{eq:criterion}
\end{equation}
which only states that, at confidence level $1-\delta$, the data do not exclude $F\geq 1/2$. In the experiments reported here, rule (iii) was used as the success criterion of the binary search, i.e., a size is accepted when the upper bound associated with the estimation error reaches or exceeds the GME fidelity threshold. The largest number of qubits satisfying this benchmark success criterion is reported as the benchmark result, $\Nst$. We stress that $\Nst$ is the largest GHZ size \emph{compatible} with GME identified by the binary-search procedure, and not the largest size for which GME is \emph{certified}; the impact of the decision rule on the results is quantified in Sec.~\ref{sec:results} (Table~\ref{tab:rules}).

\begin{figure*}[t]
    \centering
    \includegraphics{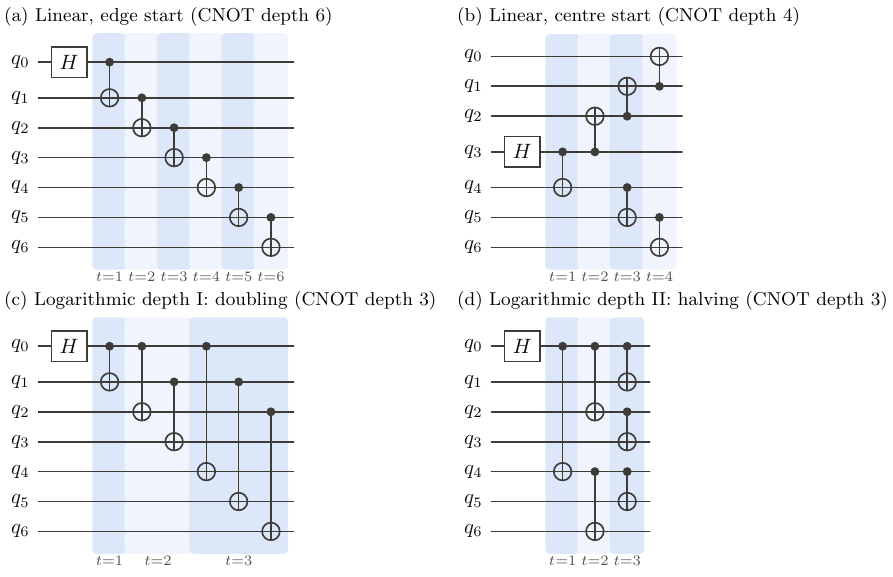}
    \caption{GHZ-state preparation circuits considered in the benchmark for $N=7$ qubits. (a) Linear strategy starting from an edge qubit. (b) Linear strategy starting from the central qubit. (c) First logarithmic-depth strategy, in which previously entangled qubits are used to incorporate additional qubits (doubling). (d) Second logarithmic-depth strategy, based on a hierarchical construction with progressively shorter-range entangling operations (halving). Shaded columns group the CNOTs of the same layer $t$, which can be executed in parallel if the connectivity allows it; the CNOT depth of (c) and (d) is $\lceil\log_2 N\rceil$.}
    \label{fig:ghz_circuits}
\end{figure*}

\subsection{GHZ circuit}

For each quantum processor and target GHZ-state size, the GHZ preparation circuit is selected by minimizing the number of native two-qubit gates after transpilation. Rather than imposing a fixed quantum circuit, four implementations of the GHZ circuit were considered: two linear strategies and two logarithmic-depth strategies, as illustrated in Fig.~\ref{fig:ghz_circuits}. The linear strategies propagate entanglement along neighboring qubits, starting either from an edge qubit or from the central qubit. In contrast, the logarithmic-depth strategies distribute entanglement hierarchically, allowing multiple entangling operations to be performed in parallel and thereby reducing circuit depth.

For each quantum processor and target GHZ-state size, all four GHZ circuits were generated and transpiled. The logical-to-physical qubit mapping was determined automatically during transpilation, and the implementation requiring the fewest native two-qubit gates after transpilation was selected for execution. The number of native two-qubit gates was used as the selection criterion because two-qubit gates contribute significantly to hardware errors. Note that, whenever the interaction graph of a strategy can be embedded in the coupling map without routing, all four strategies require exactly $N-1$ two-qubit gates, so that ties are frequent; ties were broken by circuit depth, because a shallower circuit reduces decoherence and the noise accumulated during execution. This led to the linear, centre start strategy for all QPUs except IBEX Q1, which exploits the logarithmic-depth II strategy due to its fully connected topology.

This procedure allows the GHZ preparation circuit to be adapted to the native gate set and connectivity of each quantum processor while maintaining a common circuit-selection criterion across platforms.

\subsection{Benchmark configuration}

The multipartite entanglement benchmark requires the user to specify two statistical parameters: the maximum estimation error, $\epsilon$, and the confidence parameter, $\delta$. These parameters determine the statistical precision and confidence of the fidelity estimate, and consequently the number of measurements required by the DFE protocol. Their selection therefore involves a trade-off between statistical guarantees and experimental execution cost. Throughout this work, we set the values of $\epsilon=0.1$ and $\delta=0.1$, corresponding to a confidence level $1-\delta=0.9$.

Following the benchmark definition in Ref.~\cite{aguirre2025qusquare}, the number of sampled stabilizers, $l$, is given by
\begin{equation}
    l=\left\lceil 8 \cdot \frac{\ln(4 / \delta)}{\epsilon^2}\right\rceil.
\end{equation}
For $\epsilon=0.1$ and $\delta=0.1$, this results in $l=2952$ sampled stabilizers. Each sampled stabilizer is evaluated using a single measurement shot, except for Rigetti Cepheus-1, for which AWS Braket enforces a minimum of 10 shots per circuit. In this case, only the first shot is used for fidelity estimation. Since every single-shot outcome is a $\pm1$ random variable, the standard error of $Y$ is at most $l^{-1/2}\approx 0.018$, i.e., about five times smaller than $\epsilon$.

\subsection{Experimental pipeline}

\begin{figure}[t]
    \centering
    \includegraphics{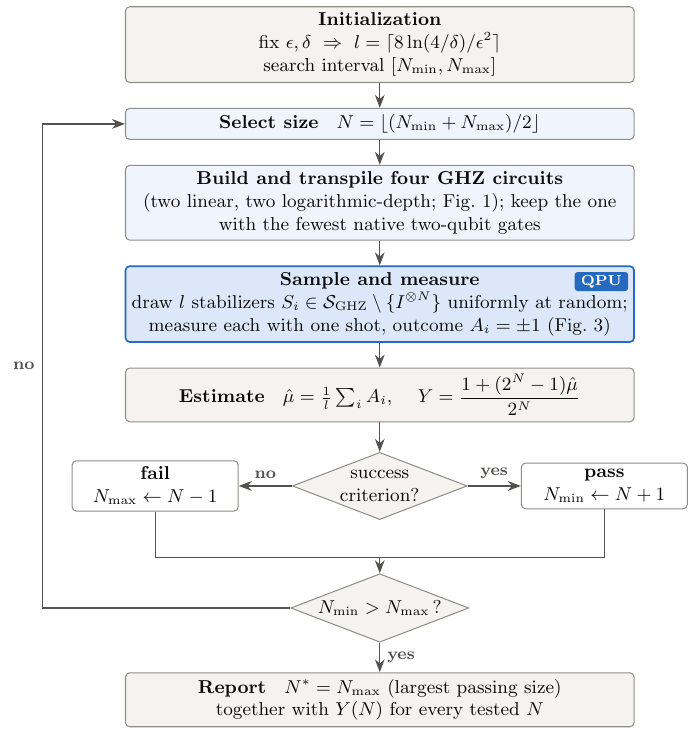}
    \caption{Experimental pipeline of the multipartite entanglement benchmark, including GHZ-circuit selection, stabilizer sampling, fidelity estimation, and the binary-search procedure used to determine the largest GHZ-state size satisfying the benchmark success criterion (Eq.~\eqref{eq:criterion}). Colours distinguish classical pre-processing (light blue), execution on the quantum processor (blue, QPU tag) and post-processing and decisions (grey).}
    \label{fig:protocol}
\end{figure}

The complete experimental pipeline is summarized in Fig.~\ref{fig:protocol}. For each target GHZ-state size selected by the binary search, the GHZ preparation circuit determined according to the procedure described above is executed on the target quantum processor.

The $l$ stabilizers are randomly sampled with uniform probability from
$\mathcal{S}_{\mathrm{GHZ}}\setminus\{I^{\otimes N}\}$ and measured on the prepared state. Each measurement yields an outcome $A_i\in\{-1,1\}$, and the sample mean is computed as
\begin{equation}
    \hat{\mu}=\frac{1}{l}\sum_{i=1}^{l}A_i.
\end{equation}
The fidelity estimator $Y$ is then obtained by accounting for the known contribution of the identity operator as
\begin{equation}
    Y = \frac{1 + (2^N - 1)\hat{\mu}}{2^N}\;=\;\hat\mu+\frac{1-\hat\mu}{2^N}.
    \label{eq:Y}
\end{equation}
All figures and tables in both the main text and the appendix report fidelity estimates computed using Eq.~\eqref{eq:Y}. These estimates are evaluated against the benchmark success criterion defined in Eq.~\eqref{eq:criterion}.

The binary search interval is device-dependent and defined as $[N_{\min},N_{\max}]=[2,\min(30,N_{\rm QPU})]$. For devices with at least 30 qubits, the upper bound was capped at 30 because our estimates indicated that no device would exceed this size while satisfying the benchmark success criterion, making further probes unnecessary. For devices with fewer than 30 qubits except Garnet, the upper bound was set to the full device size. For Garnet, the upper bound was also set to 30 by mistake, but this did not have any negative impact on the benchmark.

At each iteration, the midpoint of the current interval was tested. If the criterion is satisfied, the binary search proceeds toward larger GHZ-state sizes; otherwise, it proceeds toward smaller sizes. This procedure is repeated until the search terminates, and the largest GHZ-state size satisfying the benchmark success criterion is reported as the benchmark result. The binary search implicitly assumes that $F$ decreases monotonically with $N$. As shown in Sec.~\ref{sec:results}, this assumption is violated for some devices in single runs, so that $\Nst$ should be understood as the pass/fail boundary located by the search rather than as a strict maximum over all sizes.

While the GHZ-state fidelity provides a compact measure of benchmark performance, it does not identify the physical contributions responsible for a loss of fidelity. In particular, errors affecting the populations of the computational-basis components and errors degrading the coherence between them can lead to similar changes in the overall fidelity but have different physical origins. To provide additional diagnostic information without requiring extra measurements, we decompose the fidelity into population and coherence contributions using the same stabilizer measurement data collected by the benchmark.

\paragraph{Population--coherence decomposition.} The stabilizer group of $\ket{\mathrm{GHZ}_N}$ contains $2^{N-1}$ diagonal elements (products of an even number of Pauli-$Z$ operators) and $2^{N-1}$ off-diagonal ones (containing $X$ or $Y$ on every qubit). Averaging the two classes separately yields the population $P=\bra{0^N}\rho\ket{0^N}+\bra{1^N}\rho\ket{1^N}$ and the coherence $C=2\,\mathrm{Re}\bra{0^N}\rho\ket{1^N}$, with $F=(P+C)/2$~\cite{monz2011fourteen,Omran2019,mooney2021generation}. With uniform sampling, the corresponding estimators are
\begin{equation}
    \hat P=\frac{1+(2^{N-1}-1)\,\hat\mu_Z}{2^{N-1}},\qquad \hat C=\hat\mu_X,
    \label{eq:PC}
\end{equation}
where $\hat\mu_Z$ and $\hat\mu_X$ are the sample means over the diagonal and off-diagonal stabilizers, respectively. Since every DFE sample is labelled by the measured stabilizer, $P$ and $C$ are obtained from the same data at no additional cost, which separates population errors (bit flips, leakage and readout) from coherence errors (dephasing).


\section{Hardware}
\label{Hardware}
This benchmark was executed on nine commercially accessible quantum processing units (QPUs) spanning two physical qubit modalities and five vendors. Four backends are IBM superconducting devices from two processor families: Heron (r2 and r3) and Nighthawk (r1 and r2). One backend comes from Rigetti Computing: the chiplet-based Cepheus-1-108Q. Two backends come from IQM Quantum Computers: the 20-qubit Garnet (Crystal~20) and the 54-qubit Emerald (Crystal~54). The superconducting set is completed by the 32-qubit Qmio, built by OQC and operated by CESGA, while the trapped-ion technology is represented by the 12-qubit AQT IBEX Q1 ($^{40}$Ca$^+$). The IBM devices were accessed through the IBM Quantum Platform, Qmio through the CESGA infrastructure, and all other systems through Amazon Braket. The selection deliberately contrasts architectures with markedly different qubit-connectivity graphs, native gate sets, and scales.

\begin{figure}[t]
  \centering
  \includegraphics{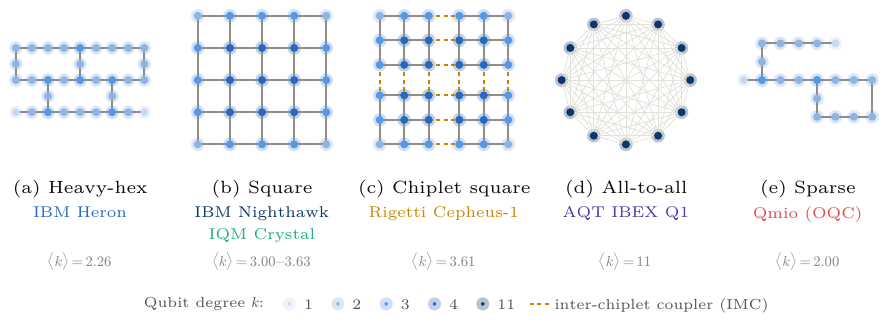}
  \caption{Schematic qubit-connectivity graphs of the five topology classes
    evaluated; node colour encodes the qubit degree $k$, and $\langle k\rangle$ is the mean degree of the actual devices (Table~\ref{tab:hardware}).
    (a)~IBM heavy-hexagonal lattice (Heron): 12-qubit rings in which interior qubits couple to
    at most three neighbours.
    (b)~Monolithic square lattice (IQM Garnet/Emerald,
    IBM Nighthawk): interior degree-four connectivity.
    (c)~Chiplet-based square lattice (Rigetti Cepheus-1-108Q, $3\times4$ array of $3\times3$ chiplets; $2\times2$ chiplets shown): solid lines are
    intra-chiplet CZ links; dashed lines are inter-chiplet intermodule
    couplers (IMCs), which carry higher error rates.
    (d)~All-to-all graph (AQT IBEX Q1, 12 ions): every qubit can directly
    interact with every other, eliminating SWAP overhead entirely.
    (e)~Sparse graph of linear chains with a few branch points (Qmio).
    Node counts and the layout of (e) are illustrative.}
  \label{fig:topologies}
\end{figure}

The most relevant technical specifications of each platform are collected in Table~\ref{tab:hardware}. Qubit counts, topology and native two-qubit gates correspond to the device specifications, while gate and readout errors, gate durations and coherence times are median values from the calibration snapshot closest to the execution of the experiments (Appendix~\ref{app:ghz}), which vary with daily calibration. Schematic connectivity graphs of the five topology classes are shown in Fig.~\ref{fig:topologies}.

\subsection{IBM Quantum Processors (Superconducting)}
\label{subsec:ibm}

IBM Quantum processors are based on fixed-frequency superconducting transmon qubits arranged in fixed planar connectivity graphs.  Single-qubit gates are implemented via microwave pulses and decomposed into $\{X, \sqrt{X}, R_z(\theta)\}$. All IBM processors evaluated here belong to the Heron and Nighthawk families~\cite{ibm-processor-types} and use a direct controlled-$Z$ (CZ) implementation enabled by tunable couplers that
suppress static qubit--qubit interaction between gates, strongly reducing crosstalk relative to the earlier Eagle generation.  Cloud access is provided through the IBM Quantum Platform~\cite{ibm-quantum-platform}. 

\subsubsection{IBM Basquecountry (Heron r2)}
\label{subsubsec:basquecountry}

\texttt{ibm\_basquecountry} is based on the Heron~r2 processor, the second revision of IBM's Heron family, released in July 2024. It integrates 156 qubits in a heavy-hexagonal lattice and incorporates a two-level-system (TLS) mitigation capability that actively manages the TLS environment of the chip, stabilising coherence across the device. 

\subsubsection{IBM Pittsburgh (Heron r3)}
\label{subsubsec:pittsburgh}

\texttt{ibm\_pittsburgh} implements the Heron~r3 architecture with 156 qubits, the third generation of IBM's Heron processor family.  Launched in July~2025 as the first publicly available Heron~r3 QPU, it retains the heavy-hexagonal layout of r2 but applies targeted fabrication improvements to coherence, gate fidelity, and readout.

\subsubsection{IBM Miami (Nighthawk r1)}
\label{subsubsec:miami}

\texttt{ibm\_miami} is based on IBM's Nighthawk~r1 architecture with 120 qubits, released in December~2025.  Nighthawk departs architecturally from
Heron: its qubits sit on a square lattice with higher interior connectivity (degree~4) relative to Heron's heavy-hexagonal lattice (degree~2 or 3). This denser graph reduces SWAP overhead. At the time of the experiments, \texttt{ibm\_miami} had exploratory status and was still being calibrated. Its median readout error (1.99\%) stands out as the highest among the IBM devices. 

\subsubsection{IBM Phoenix (Nighthawk r2)}
\label{subsubsec:phoenix}

\texttt{ibm\_phoenix} runs the Nighthawk~r2 processor, released on September~2026.  It retains the 120-qubit square-lattice architecture of
Nighthawk~r1 (\texttt{ibm\_miami}) but replaces conditional reset with a dissipative reset gadget: each programmable qubit is coupled through a tunable coupler to a cold environment that returns it to its ground state on demand, cutting the default repetition delay between circuit executions. IBM reports roughly $25\times$ higher circuit throughput than the Heron fleet as a result, while maintaining Heron-class gate fidelities. Its median coherence times, however, are lower than those of the Heron devices. The \texttt{ibm\_phoenix} experiments were performed on 3~September~2026, seven weeks after those on the other IBM devices. 

\subsection{Rigetti Cepheus-1-108Q (Superconducting, 108 Qubits)}
\label{subsec:rigetti}
Rigetti Cepheus-1-108Q is a 108-qubit superconducting processor built from
twelve 9-qubit chiplets tiled in a $3\times4$ array~\cite{rigetti_qpus}.
Within each chiplet, transmon qubits form a square lattice connected by tunable
couplers, and intermodule couplers (IMCs) link neighbouring chiplets. Each qubit
couples to up to four nearest neighbours. Segmenting the processor allows the
chiplets to be fabricated, tested and optimised independently, decoupling
fabrication complexity from total qubit count. The price is a class of
inter-chiplet links with no counterpart in monolithic processors. The native
two-qubit gate is an adiabatic CZ, an implementation that increases resilience
to leakage errors~\cite{braket-rigetti}.


\subsection{IQM Quantum Processors (Superconducting)}
\label{subsec:iqm}
IQM quantum processors are based on superconducting transmon qubits arranged in
a square lattice and connected by tunable couplers (IQM Crystal architecture).
The native single-qubit gate is the phased-$X$ rotation (PRX), which implements
arbitrary $X$ and $Y$ rotations, and the native two-qubit gate is the CZ~\cite{braket-iqm}. The two devices evaluated here share this
gate set and topology class, so that the architectural differences between them are limited to scale. 

\subsubsection{IQM Garnet (Crystal 20)}
\label{subsubsec:garnet}

IQM Garnet is a 20-qubit processor containing, in addition to the
computational qubits, 30 tunable coupler qubits that define its square lattice connectivity~\cite{iqm_garnet}.

\subsubsection{IQM Emerald (Crystal 54)}
\label{subsubsec:emerald}

IQM Emerald implements the Crystal architecture with 54 qubits~\cite{iqm_emerald}. It extends the square-lattice architecture of Garnet, while exhibiting lower median gate errors and longer coherence times at the time of our experiments (Table~\ref{tab:hardware}).

\subsection{AQT IBEX Q1 (Trapped Ion)}
\label{subsec:aqt}

The AQT IBEX Q1 is a 12-qubit trapped-ion quantum computer operated by Alpine Quantum Technologies (AQT)~\cite{aqt_ibex}.  It traps a linear chain of $^{40}$Ca$^+$ ions in a radio-frequency Paul trap and implements single-qubit rotations via addressed laser pulses and two-qubit M\o{}lmer--S\o{}rensen gates mediated by the shared motional modes of the ion crystal.  The all-to-all connectivity provided by the shared phonon bus allows any pair of qubits to interact directly without SWAP overhead.  The native gate set comprises RZ, R (arbitrary single-qubit rotation), and RXX operations.

Its defining engineering feature is form factor: the complete system fits in two standard 19-inch telecom racks, operates at room temperature (no dilution refrigerator required), and consumes less than \SI{2}{\kilo\watt}, which makes it suitable for deployment in standard data-centre environments. 


\subsection{Qmio (Superconducting)}
\label{subsec:qmio}

Qmio is a 32-qubit superconducting processor with a coaxmon qubit architecture, built by OQC and deployed on-premises at CESGA as part of a tightly integrated HPC--QC system~\cite{cacheiro2025qmio}.  Unlike the regular 2D lattices used by the other superconducting platforms in this study, Qmio's coupling graph is a sparse network of linear qubit chains punctuated by a handful of degree-3 branch points, with a mean degree of 2.0.

\subsection{Comparative summary}
\label{subsec:summary}
\begin{table*}[ht!]
  \centering
  \caption{Hardware parameters of the evaluated QPUs. Qubit counts, topology, and the native two-qubit gate (2Q) are device specifications, where $\langle k\rangle$ is the mean qubit degree of the coupling graph. Single- and two-qubit error rates (1Q err., 2Q err.) are per-gate error probabilities (not fidelities), given in percent, and RO err.\ is the readout error. Error rates and coherence times ($T_1$, $T_2$) are median values extracted from the device calibration data at the time of experiment execution (Appendix~\ref{app:ghz}). A dash (--) indicates a value not reported by the provider or not recorded. Values marked with $^\ast$ were taken from the vendor's documentation rather than obtained from the experiments. \texttt{ibm\_miami} held exploratory status and was still being calibrated during the experiments. Coherence times of the trapped-ion system (AQT IBEX Q1) are orders of magnitude longer than those of the superconducting devices.}
  \label{tab:hardware}
  \scriptsize
  \setlength{\tabcolsep}{3pt}
  \renewcommand{\arraystretch}{1.2}
  \begin{tabular}{@{}l r l c c c c c c c@{}}
    \toprule
    \textbf{Backend} (processor) & \textbf{Qubits} & \textbf{Topology} & $\boldsymbol{\langle k\rangle}$ & \textbf{2Q} &
    \textbf{1Q err.} & \textbf{2Q err.} & \textbf{RO err.} & \textbf{Gate time} & \textbf{$T_1$\,/\,$T_2$} \\
     & & & & & (\%) & (\%) & (\%) & \textbf{1Q/2Q} & \\
    \midrule
    ibm\_basquecountry (Heron r2) & 156 & Heavy-hex & 2.26 & CZ
      & 0.036 & 0.183 & 0.65 & 32/68 ns & 253/205\,$\mu$s \\
    ibm\_pittsburgh (Heron r3) & 156 & Heavy-hex & 2.26 & CZ
      & 0.019 & 0.157 & 0.42 & 32/88 ns & 285/312\,$\mu$s \\
    ibm\_miami (Nighthawk r1) & 120 & Square & 3.63 & CZ
      & 0.024 & 0.319 & 1.99 & 32/138 ns & 304/205\,$\mu$s \\
    ibm\_phoenix (Nighthawk r2) & 120 & Square & 3.63 & CZ
      & 0.022 & 0.203 & 0.73 & 40/68 ns & 176/140\,$\mu$s \\
    Rigetti Cepheus-1-108Q & 108 & Chiplets & 3.61 & CZ
      & 0.263 & 1.312 & 3.10 & 40/60 ns$^\ast$ & 25/9\,$\mu$s \\
    IQM Garnet & 20 & Square & 3.00 & CZ
      & 0.086 & 0.670 & 1.51 & -- & 32/9 \,$\mu$s \\
    IQM Emerald & 54 & Square & 3.15 & CZ
      & 0.034 & 0.418 & 1.24 & -- & 52/16 \,$\mu$s \\
    AQT IBEX Q1 & 12 & All-to-all & 11.00 & XX
      & 0.029 & 1.570 & 0.22 & 45/335 $\mu$s & 1\,s/163\,ms \\
    Qmio (OQC) & 32 & Sparse & 2.00 & ECR
      & 0.433 & 6.411 & 12.56 & 30/150 $\mu$s & 54/77\,$\mu$s \\
    \bottomrule
  \end{tabular}
\end{table*}


\section{Software}
\label{Software}

The experimental pipeline and benchmarking framework were implemented using the Qiskit and Amazon Braket software development kits (SDKs); Qiskit was also used for Qmio, which was accessed through the CESGA infrastructure~\cite{cacheiro2025qmio}. To ensure transparency and reproducibility, the complete codebase, including circuit generation, optimization, execution, and data post-processing routines, is publicly available, as described in Sec.~\ref{sec:data}. 

Due to differences between the software stacks and execution workflows of the considered platforms, platform-specific compilation and transpilation configurations were required. For quantum processors accessed through Qiskit, circuits were transpiled using \texttt{optimization\_level=3} to optimize the circuits and reduce gate overhead prior to execution. For processors accessed through Amazon Braket, the final compilation is managed by the hardware provider as part of the default cloud execution workflow. Unlike Qiskit, where optimization levels can be explicitly configured, the compilation rules applied by the hardware provider are not directly configurable by the user through Amazon Braket. Amazon Braket does, however, offer a verbatim-compilation mode for the AQT, IQM and Rigetti devices, in which native-gate circuits are executed without modification on user-selected physical qubits~\cite{braket-verbatim}. It was deliberately not used here, so that the results reflect the default user experience. An important consequence is an asymmetry between toolchains: at optimization level 3 the Qiskit transpiler selects the physical layout using the current calibration data (noise-aware layout), whereas the default provider compilation on Braket does not necessarily avoid poorly performing qubits~\cite{tannu2019notall}. This asymmetry is part of the system under test and must be kept in mind when comparing vendors.

All error mitigation techniques were disabled to evaluate the processors without additional error-suppression or post-processing methods. For executions through Qiskit, Pauli twirling and error resilience features were explicitly disabled, and dynamical decoupling was not enabled. For Amazon Braket executions, the default provider configurations were used without additional error mitigation.

\begin{figure}[t]
    \centering
    \includegraphics{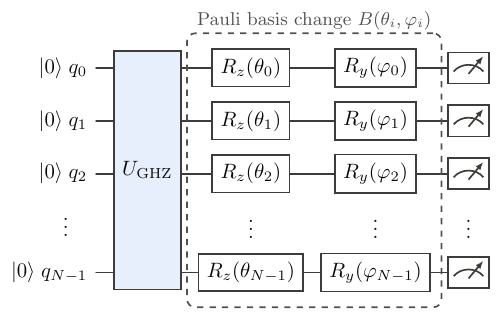}
    \caption{Parameterized circuit used to implement the Pauli measurement bases. $U_{\mathrm{GHZ}}$ denotes the circuit used to prepare the GHZ state. Single-qubit rotations $R_z(\theta_i)$ and $R_y(\varphi_i)$ are applied after GHZ-state preparation and before measurement; they map the Pauli operator of the sampled stabilizer on qubit $i$ onto the computational basis, with $(\theta_i,\varphi_i)=(0,0)$ for $Z$, $(0,-\pi/2)$ for $X$ and $(-\pi/2,-\pi/2)$ for $Y$. The sign of the stabilizer is applied in post-processing}.
    \label{fig:parametrized_circuit}
\end{figure}

The DFE protocol requires measurements of multiple Pauli observables. Executing each measurement basis as a separate circuit substantially increases execution time and cost. However, on IBM Quantum, submitting multiple commuting Pauli observables using any version of the Estimator primitive results in only one observable being executed to reduce costs. This optimization is incompatible with the measurement procedure required by the DFE protocol, since it considerably reduces the real shots and, therefore, the confidence level of the experiment. To address this limitation, we implemented the parameterized measurement scheme illustrated in Fig.~\ref{fig:parametrized_circuit}. The circuit $U_{\mathrm{GHZ}}$ prepares the GHZ state, after which parameterized single-qubit rotations $R_z(\theta_i)$ and $R_y(\varphi_i)$ are applied to each qubit $i$ before measurement. By adjusting the rotation angles, the measurement basis of each qubit can be changed according to the Pauli observable being evaluated. Using the IBM Sampler primitive, the different measurement configurations were executed by assigning different rotation angles to the same parameterized transpiled circuit. This approach avoided the observable reduction described above while reducing transpilation and execution overhead.

For IBM and Qmio, circuit transpilation was first performed locally to compare the four GHZ preparation strategies and identify the circuit requiring the fewest native two-qubit gates. Since Amazon Braket does not support local transpilation and most of its QPUs share topologies similar to those analyzed on IBM, the same preparation strategy was applied. The only exception was IBEX Q1, which used a different strategy because its fully connected topology allowed a logarithmic-depth construction without introducing SWAP gates. The selected circuit was then submitted to the corresponding Amazon Braket backend, where the final compilation was handled by the hardware provider as part of the cloud execution workflow. 

Additional platform- and QPU-specific constraints required modifications to the execution strategy. On Amazon Braket, circuit configurations were grouped into program sets to reduce execution overhead and cost. A batch size of 100 elements was used for most devices. For the AQT trapped-ion system, result retrieval was limited to the first 50 elements, and the batch size was therefore reduced accordingly.

It is worth noting that Rigetti Cepheus-1 requires a minimum of 10 shots per circuit. However, to maintain consistency, the extra runs were not taken into account to compute the metrics.

Executions on the Qmio quantum computer were also subject to hardware-level execution time limits. During the benchmark, server-side middleware timeouts were encountered, with a limit of $25{,}000\,\mathrm{ms}$ reported through a \texttt{QSLAMException}. The Qmio experiments were therefore divided into multiple runs to ensure successful execution.

\section{Results}
\label{sec:results}

\begin{figure}[htbp]
    \centering
    \includegraphics{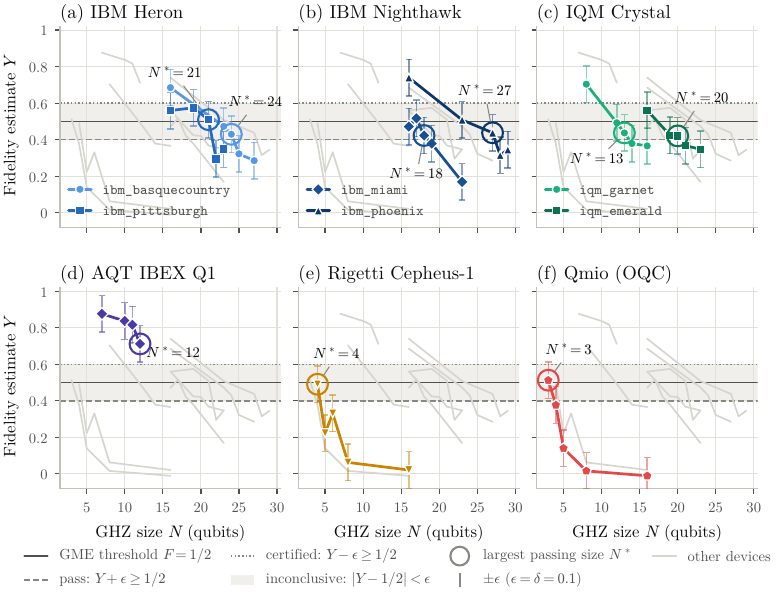}
    \caption{Fidelity estimate $Y$ vs.\ the number of qubits used to generate the GHZ state. Panels group the processors by family; in each panel the other devices are shown in grey. Colours encode the vendor (blue: IBM, green: IQM, violet: AQT, amber: Rigetti, red: Qmio), and markers distinguish devices of the same vendor. Error bars denote uncertainty determined by the benchmark parameters $\epsilon=0.1$ and $\delta=0.1$; the standard error of each point ($\leq0.018$) is comparable to the marker size.  Circles mark the largest passing size $\Nst$. Solid line: GME threshold $F=1/2$; dashed line: success criterion of Eq.~\eqref{eq:criterion}, $Y+\epsilon\geq1/2$; dotted line: certification threshold $Y-\epsilon\geq1/2$. Values are computed from the per-size estimates listed in Appendix~\ref{app:ghz}.
    }
    \label{fig:fidelity_vs_nqubits}
\end{figure}

In Fig.~\ref{fig:fidelity_vs_nqubits}, we present the GHZ state fidelity estimate $Y$ (Eq.~\eqref{eq:Y}) as a function of the number of qubits $N$ across all evaluated quantum processors, grouped by processor family; each panel highlights one family and shows the remaining devices in grey for reference. The error bars represent the statistical uncertainty bounds set by our benchmark parameter selection ($\epsilon = 0.1$ and $\delta = 0.1$). The metric of the benchmark, marked by circles, corresponds to the largest number of qubits $\Nst$ satisfying Eq.~\eqref{eq:criterion}, i.e., with $Y\geq 0.4$ (dashed line). The dotted line at $Y=0.6$ marks the certification threshold $Y-\epsilon\geq 1/2$; points in the shaded band between both lines are statistically inconclusive with respect to GME.

As anticipated for noisy intermediate-scale quantum (NISQ) devices, state fidelity decreases overall with increasing qubit count due to error accumulation in larger circuits. The $\pm\epsilon$ bars are, however, a conservative worst-case bound, whereas the standard error of each point is at most $l^{-1/2}\approx0.018$ (Sec.~\ref{methodology}). On this scale, the small inversions observed for \texttt{ibm\_pittsburgh} ($N=16\to19$) and \texttt{ibm\_miami} ($N=16\to17$) are within about two standard errors, but the drop of \texttt{ibm\_pittsburgh} between $N=21$ ($Y=0.51$) and $N=22$ ($Y=0.30$) and the inversion of \texttt{rigetti\_cepheus} between $N=5$ ($Y=0.22$) and $N=6$ ($Y=0.33$) exceed four standard errors. Since each size was measured once, with an independently transpiled circuit, these features cannot be attributed to shot noise; they reflect changes in the physical qubits selected by the compiler and/or calibration drift between runs.

The superconducting architectures from IBM achieved the highest absolute qubit capacity meeting the benchmark criteria: \texttt{ibm\_phoenix} (Nighthawk r2) led with $N^*=27$, followed by \texttt{ibm\_basquecountry} (Heron r2) at $N^*=24$, and \texttt{ibm\_pittsburgh} (Heron r3) at $N^*=21$. These were followed by \texttt{iqm\_emerald} (superconducting) with $N^*=20$ and \texttt{ibm\_miami} (Nighthawk r1) with $N^*=18$. The remaining platforms recorded lower maximum capacities: \texttt{iqm\_garnet} ($N^*=13$), the trapped-ion processor \texttt{aqt\_ibex\_q1} ($N^*=12$), and the superconducting processors \texttt{rigetti\_cepheus} ($N^*=4$) and \texttt{qmio} ($N^*=3$). Notably, \texttt{aqt\_ibex\_q1} delivered the highest fidelity ($\approx 0.71$) at its largest tested size, which coincides with its full register ($N=12$).

The size at which a device passes the benchmark should not be confused with the largest size at which GME is certified. Table~\ref{tab:rules} lists, for every device, the pass/fail boundary obtained with each of the three decision rules of Sec.~\ref{methodology}. Because only the sizes visited by the binary search were measured, the boundaries for rules (i) and (ii) are in general intervals. Under the certification rule, only four devices certify GME at any of the tested sizes (\texttt{ibm\_phoenix} and \texttt{ibm\_basquecountry} at $N=16$, \texttt{iqm\_garnet} at $N=8$ and \texttt{aqt\_ibex\_q1} at $N=12$), and the ordering changes: \texttt{ibm\_pittsburgh}, which passes the benchmark up to $N=21$, does not certify GME at any tested size.

\begin{table*}[t]
\centering
\caption{Summary of the benchmark results and sensitivity of the pass/fail boundary to the decision rule. $N_{\rm tot}$: number of available qubits; $\Nst$: result of the benchmark as executed [rule (iii), $Y+\epsilon\geq1/2$]; $\varepsilon_q=1-Y(\Nst)^{1/\Nst}$: per-qubit GHZ error. The last two columns give the boundary obtained with rule (ii) ($Y\geq1/2$) and rule (i) ($Y-\epsilon\geq1/2$, certification); ``$a$--$b$'' means that size $a$ passes and the next tested size, $b+1$, fails, and ``$<a$'' that no tested size passes. $^{\dagger}$Non-monotonic: $Y(16)<1/2<Y(17)$. $^{\ddagger}$Limited by the register size (lower bound).}
\label{tab:rules}
\footnotesize
\setlength{\tabcolsep}{3.5pt}
\renewcommand{\arraystretch}{1.12}
\begin{tabular}{@{}l c l l c c c c c c c c@{}}
\toprule
Device & $N_{\rm tot}$ & Tested sizes & Date (2026)& $\Nst$ & $\Nst/N_{\rm tot}$ & $Y(\Nst)$ & $\varepsilon_q$ & $Y\geq\frac12$ & $Y-\epsilon\geq\frac12$ & $\hat P$& $\hat C$ \\
\midrule
\texttt{ibm\_phoenix}        & 120 & 16, 23, 27, 28, 29 & 09-03 & 27 & 23\%  & 0.438 & 3.0\%  & 23--26 & 16--22 & 0.66 & 0.21\\
\texttt{ibm\_basquecountry} & 156 & 16, 23, 24, 25, 27 & 07-14 & 24 & 15\%  & 0.430 & 3.5\%  & 16--22 & 16--22 & 0.69 & 0.16\\
\texttt{ibm\_pittsburgh}    & 156 & 16, 19, 21, 22, 23 & 07-14 & 21 & 13\%  & 0.510 & 3.2\%  & 21     & $<16$ & 0.73 & 0.29\\
\texttt{iqm\_emerald}       & 54  & 16, 19, 20, 21, 23 & 07-14 & 20 & 37\%  & 0.421 & 4.2\%  & 16--18 & $<16$ & 0.54 & 0.31 \\
\texttt{ibm\_miami}         & 120 & 16, 17, 18, 19, 23 & 07-14 & 18 & 15\%  & 0.423 & 4.7\%  & 17$^{\dagger}$ & $<16$ & 0.63 & 0.21\\
\texttt{iqm\_garnet}        & 20  & 8, 12, 13, 14, 16  & 07-14 & 13 & 65\%  & 0.437 & 6.2\%  & 8--11  & 8--11 & 0.51 & 0.36\\
\texttt{aqt\_ibex\_q1}      & 12  & 7, 10, 11, 12      & 07-27 & 12$^{\ddagger}$ & 100\% & 0.711 & 2.8\%  & 12$^{\ddagger}$ & 12$^{\ddagger}$ & 0.74 & 0.68 \\
\texttt{rigetti\_cepheus}   & 107 & 4, 5, 6, 8, 16     & 07-11 & 4  & 4\%   & 0.491 & 16.3\% & $<4$   & $<4$ & 0.69 & 0.29 \\
\texttt{qmio}               & 32  & 3, 4, 5, 8, 16     & 07-22/27 & 3 & 9\% & 0.513 & 20.0\% & 3      & $<3$ & 0.67 & 0.35 \\
\bottomrule
\end{tabular}
\end{table*}

\begin{figure}[htbp]
    \centering
    \includegraphics{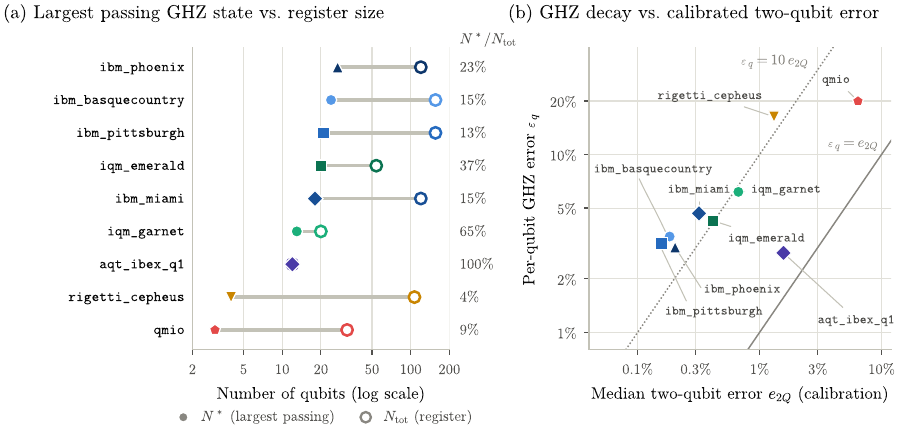}
    \caption{(a) Largest passing GHZ size $\Nst$ (filled markers) compared with the register size $N_{\rm tot}$ (open circles), on a logarithmic scale; the ratio $\Nst/N_{\rm tot}$ is given on the right. For \texttt{aqt\_ibex\_q1}, $\Nst=N_{\rm tot}$ and the result is limited by the register size. (b) Per-qubit GHZ error $\varepsilon_q=1-Y(\Nst)^{1/\Nst}$ versus the median two-qubit gate error of the calibration snapshot (Table~\ref{tab:hardware}). Solid line: $\varepsilon_q=e_{2Q}$; dotted line: $\varepsilon_q=10\,e_{2Q}$.}
    \label{fig:summary}
\end{figure}

\begin{figure}[htbp]
    \centering
    \includegraphics{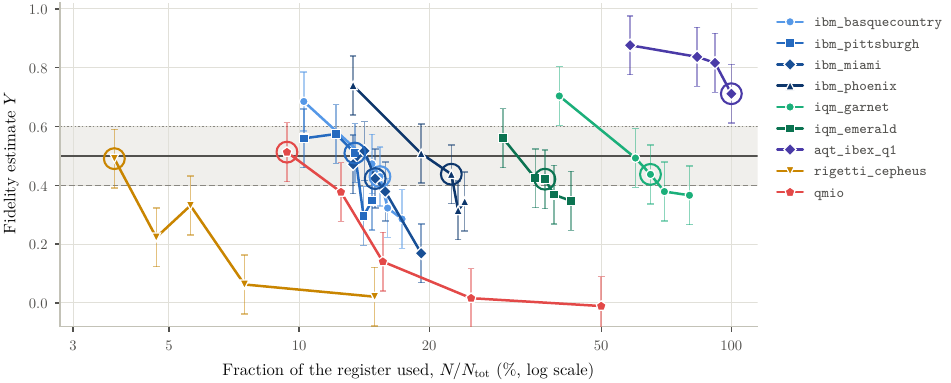}
    \caption{Fidelity vs. the percentage of total QPU capacity utilized to construct the GHZ state. Colors denote distinct quantum processors, with error bars reflecting the uncertainty bounds ($\epsilon=0.1$, $\delta=0.1$). Circles highlight the maximum QPU percentage satisfying the benchmark threshold; lines and shading as in Fig.~\ref{fig:fidelity_vs_nqubits}.}
    \label{fig:fidelity_vs_qpu-percentage}
\end{figure}

Figure~\ref{fig:summary}(a) compares $\Nst$ with the register size, and Fig.~\ref{fig:fidelity_vs_qpu-percentage} evaluates fidelity as a function of the QPU utilization percentage, defined as the fraction of the processor's total available qubit register used to construct the GHZ state. This metric significantly shifts platform rankings. The trapped-ion processor \texttt{aqt\_ibex\_q1} ranks highest, reaching $100\%$ utilization while satisfying the benchmark metric, followed by \texttt{iqm\_garnet} ($65\%$) and \texttt{iqm\_emerald} ($37\%$). IBM processors rank lower on relative scale due to their larger overall registers: \texttt{ibm\_phoenix} ($23\%$), \texttt{ibm\_basquecountry} ($15\%$), \texttt{ibm\_miami} ($15\%$), and \texttt{ibm\_pittsburgh} ($13\%$). Finally, \texttt{qmio} and \texttt{rigetti\_cepheus} yield maximum utilization values of $9\%$ and $4\%$, respectively.

This relative metric must be interpreted with care: it is bounded by 100\%, it is trivially maximized by small registers (the result of \texttt{aqt\_ibex\_q1} only provides the lower bound $\Nst\geq12$), and it penalizes large processors even when they generate larger GHZ states. We therefore complement it with a size-independent measure, the per-qubit GHZ error $\varepsilon_q\equiv1-Y(\Nst)^{1/\Nst}$, i.e., the constant infidelity per qubit that would reproduce $Y(\Nst)$ if every additional qubit contributed the same multiplicative factor. As shown in Table~\ref{tab:rules} and Fig.~\ref{fig:summary}(b), $\varepsilon_q$ ranges from 2.8\% (\texttt{aqt\_ibex\_q1}) and 3.0\% (\texttt{ibm\_phoenix}) to 16\% (\texttt{rigetti\_cepheus}) and 20\% (\texttt{qmio}). For all superconducting devices except Qmio, $\varepsilon_q$ exceeds the median two-qubit gate error by a factor of about 9--20, whereas for \texttt{aqt\_ibex\_q1} the ratio is only $\approx1.8$.

\section{Discussion}
\label{sec:discussion}

The high fidelity performance achieved by \texttt{aqt\_ibex\_q1} aligns with expectations for trapped-ion systems. All-to-all connectivity allows direct multi-qubit entanglement without routing overhead and, more importantly, provides greater flexibility in selecting the physical qubits and couplings used by a circuit. For circuits that do not exceed the threshold, the required qubit chains fit within the available QPU topologies, so no additional SWAP gates are required. The advantage of higher connectivity therefore comes not from reducing the number of two-qubit gates, but from providing a larger set of possible qubit mappings, allowing the transpiler to select more favorable qubits and connections. Therefore, although this QPU's total physical capacity ($N=12$) is lower compared to modern superconducting chips, \texttt{aqt\_ibex\_q1} meets the benchmark across its entire hardware register, leading to top performance in relative QPU utilization ($100\%$). Two caveats apply, however. First, its $\Nst$ is limited by the register size and not by the fidelity ($Y=0.71$ at $N=12$), so the benchmark only provides a lower bound. Second, its per-qubit GHZ error (2.8\%) is comparable to that of the best superconducting devices (3.0\% for \texttt{ibm\_phoenix}), and \texttt{ibm\_phoenix} reached a higher fidelity at $N=16$ ($Y=0.74$) than \texttt{aqt\_ibex\_q1} at $N=12$ ($Y=0.71$). The advantage of the trapped-ion device in this benchmark thus stems from its small register rather than from a superior per-qubit quality. Interestingly, its median two-qubit error (1.6\%) is the second largest among all devices, but its per-qubit GHZ error is only $\approx1.8$ times larger (Fig.~\ref{fig:summary}(b)), consistent with a GHZ decay dominated by the entangling gates themselves.

IBM superconducting processors achieved the largest absolute qubit scale meeting the benchmark threshold ($Y+\epsilon\geq 1/2$). This performance correlates with their low two-qubit (2Q) error rates ($1.6\times10^{-3}$ to $3.2\times10^{-3}$), which are the lowest among all evaluated superconducting systems. However, two-qubit gate errors alone cannot account for the observed decay: at $N=24$, the 23 CZ gates of \texttt{ibm\_basquecountry} contribute only $\approx4\%$ infidelity, whereas the measured infidelity is $\approx57\%$. Readout errors, which affect all qubits in every stabilizer measurement and were not mitigated (0.4--2\% per qubit), the idling errors of qubits waiting to be entangled in linear-depth circuits, and crosstalk are expected to dominate, as reflected by per-qubit GHZ errors 15--20 times larger than the median two-qubit error (Fig.~\ref{fig:summary}(b)). Among IBM devices, \texttt{ibm\_phoenix} demonstrated the best scaling ($N^*=27$), benefiting from a square grid topology which provides greater flexibility in qubit and coupling selection than heavy-hex layouts. Note, however, that \texttt{ibm\_phoenix} was measured seven weeks later than the other IBM devices and that its median coherence times are shorter than those of the Heron processors (Table~\ref{tab:hardware}); the topological argument is consistent with the observation but does not prove it. Although \texttt{ibm\_miami} also features a square topology, its performance ($N^*=18$) was limited by higher native 2Q gate errors, a median readout error three times larger than on the other IBM devices, and its  exploratory status at execution time. The heavy-hex devices \texttt{ibm\_basquecountry} (Heron r2) and \texttt{ibm\_pittsburgh} (Heron r3) reached similar results ($N^*=24$ and $N^*=21$, respectively). Remarkably, \texttt{ibm\_pittsburgh} has better median calibration figures than \texttt{ibm\_basquecountry} (lower one-qubit, two-qubit and readout errors, and longer $T_2$) but a smaller $\Nst$, owing to the abrupt fidelity drop between $N=21$ and $N=22$. This illustrates that device-wide medians are poor predictors of the quality of the specific qubit chain selected in a given run~\cite{tannu2019notall}. Under the relative QPU metric, IBM processors rank lower than \texttt{aqt\_ibex\_q1} and IQM systems because their large physical chip capacities dilute percentage scaling despite superior absolute qubit counts.

The IQM superconducting processors, \texttt{iqm\_emerald} and \texttt{iqm\_garnet}, bridge the performance gap between IBM architectures and lower-performing systems (\texttt{qmio} and \texttt{rigetti\_cepheus}). This performance reflects their intermediate 2Q gate error rates and shorter coherence times ($T_2=9$--$16\,\mu$s). Between the two, \texttt{iqm\_emerald} features lower 2Q error rates and correspondingly outperforms \texttt{iqm\_garnet} both in $\Nst$ and in per-qubit GHZ error (4.2\% vs.\ 6.2\%). On the relative utilization metric, IQM machines outperform IBM systems due to their smaller total physical registers ($N_{\text{tot}} = 54$ for \texttt{iqm\_emerald} and $N_{\text{tot}} = 20$ for \texttt{iqm\_garnet}). Recently, IQM issued a press release \cite{iqm_ghz_2026} claiming the preparation of a 54-qubit GHZ state with a fidelity exceeding 0.5. Two key differences distinguish their results from our computations: first, IQM applied error mitigation techniques; second, our calculations were executed via cloud access on Amazon Braket. 

Conversely, \texttt{rigetti\_cepheus} and \texttt{qmio} exhibited the lowest results, $\Nst=4$ and $\Nst=3$, respectively, both in absolute and in relative terms. For \texttt{qmio}, this is consistent with median ECR and single-qubit errors (6.4\% and 0.43\%), the largest among all evaluated devices, and with a sparse coupling graph ($\langle k\rangle=2.0$) that restricts the choice of qubit chains. The case of \texttt{rigetti\_cepheus} is different: its coupling graph is a dense square lattice ($\langle k\rangle=3.61$, comparable to IBM Nighthawk) and its median CZ error (1.3\%) is similar to that of \texttt{aqt\_ibex\_q1}, yet its fidelity collapses beyond $N=4$ ($Y=0.22$ at $N=5$ and $0.06$ at $N=8$), with the second largest per-qubit GHZ error (16\%). This behaviour points to the inclusion of one or more faulty qubits or couplers, possibly aggravated by its short coherence times ($T_2=9.3\,\mu$s) and by the intermodule couplers. Additionally, in the case of \texttt{rigetti\_cepheus}, the default provider compiler does not explicitly select qubits based on their calibration error rates, and consequently selected a faulty qubit in our experiments by chance. Because this benchmark evaluates not only the underlying QPU but the complete execution platform, we treat the  default compilation workflow as part of the system under evaluation. Nevertheless, to investigate the impact of the automatic qubit-selection strategy on the observed performance, Appendix~\ref{ceph} reports results obtained using manually selected physical qubits.


\subsection{Limitations and outlook}
\label{subsec:limitations}

Several aspects of the present study limit the generality of its conclusions and point to natural extensions.
\begin{itemize}
    \item \emph{Single runs and temporal variability.} Each size was measured once, at a single date per device (Table~\ref{tab:rules}).  In principle, repeating the benchmark several times, and on different days, would allow reporting a distribution of $\Nst$ rather than a single value~\cite{teegarden2026three}, reducing the dependence on calibration. To partially assess the impact of temporal variability, we repeated the benchmark on three IBM Quantum processors at different points in time. While individual fidelity estimates varied between executions, the maximum GHZ-state size identified by the benchmark remained largely stable. A detailed analysis is provided in Appendix~\ref{temp_var}.
    
    \item \emph{Binary search.} The search assumes a monotonic decrease of the fidelity with $N$, which is violated in single runs (Sec.~\ref{sec:results}). Measuring all sizes in a window around the boundary would make $\Nst$ more robust. However, budget restrictions have limited this analysis. 
    \item \emph{Toolchain asymmetry.} The noise-aware layout of Qiskit and the default provider compilation of Amazon Braket are not equivalent. However, this is a limitation that an average user would find. 
    \item \emph{No error mitigation.} Readout errors enter every stabilizer measurement and grow with $N$. A metric with readout-error mitigation (e.g., M3~\cite{nation2021scalable}) would disentangle state preparation from measurement. The 120-qubit GHZ state of Ref.~\cite{javadiabhari2025bigcats} illustrates the gap between the default user experience measured here and expert-level optimization. However, fairly implementing them in such a variety of processors and accesses is highly complicated. Additionally, it would increase the ``value'' of the metric, but expectedly not the relative behavior.
    \item \emph{Technology coverage.} The trapped-ion side of the comparison rests on a single 12-qubit, register-limited device due to the extremely high cost of running them in larger trapped-ion. We are currently trying to address this problem.
\end{itemize}

\section{Conclusion}
\label{Conclusions}
We have benchmarked nine cloud-accessible quantum processors from five vendors, based on superconducting and trapped-ion technologies, using the multipartite entanglement benchmark of the QuSquare suite. The fidelity of GHZ states was estimated by direct fidelity estimation, with a number of samples independent of the number of qubits, using the default compilation of each provider and without error mitigation, so that the results reflect the performance experienced by end users.

The largest GHZ states satisfying the benchmark criterion range from $\Nst=3$ (Qmio) and $\Nst=4$ (Rigetti Cepheus-1) to $\Nst=27$ (IBM Nighthawk r2), and remain below a quarter of the register on the superconducting processors with more than 100 qubits, whereas the 12-qubit trapped-ion processor passes with its full register. Normalizing by the register size therefore favours small processors, while the per-qubit GHZ error provides a size-independent comparison in which the best superconducting and trapped-ion devices are comparable ($\approx3\%$). On the superconducting processors, the per-qubit GHZ error exceeds the median two-qubit gate error by about one order of magnitude, indicating that readout, idling and crosstalk errors, together with the quality of the qubit selection, dominate the performance that a user obtains through the cloud. We have also shown that the benchmark outcome depends on the statistical decision rule: requiring certified genuine multipartite entanglement, $Y-\epsilon\geq1/2$, only four of the nine processors succeed at any of the tested sizes.

Future work will repeat the measurements over time to quantify temporal variability, adopt the certification criterion throughout, include readout-mitigated and verbatim-compiled variants to separate hardware from software effects, and extend the comparison to larger trapped-ion and neutral-atom processors and to dynamic-circuit GHZ preparation.

\section*{Data availability}
\label{sec:data}
The data that support the results of this study are openly available at Zenodo~\cite{hernandez2026ghzdataset}
under a CC BY 4.0 licence. The dataset contains the results, the calibration snapshot of each processor at execution time, and the ideal and executed circuits together with the physical qubits on which they run. The code that generated and processed this data is available in Ref~\cite{nquire2026benchmarkingcode}.

\section*{Acknowledgments}

This research was funded by BIQAIN initiative of the Bizkaiko Foru Aldundia through Telefonica services. The authors also thankfully acknowledge support from Ikerbasque and BasQ for the access to the \texttt{ibm\_basquecountry} processor, from Amazon Braket under the AWS Cloud Credit for Research program for the access to IQM, Rigetti, and AQT quantum processors. The authors thankfully acknowledge RES resources provided by CESGA in Finisterrae-QMIO-QPU to FI-2026-1-0033.

D.A., R.P. and M.S. acknowledge support from OpenSuperQ+100 (Grant No. 101113946) of the EU Flagship on Quantum Technologies, from Project Grant No. PID2024-156808NB-I00 and Spanish Ramón y Cajal Grant No. RYC-2020-030503-I funded by MICIU/AEI/10.13039/501100011033 and by ``ERDF A way of making Europe'' and ``ERDF Invest in your Future'', and from Basque Government grant IT1887-26 and the Q-STREAM II project of the BasQ initiative. This work has additionally received funding from the European Union's Horizon Europe research and innovation programme through the JANUS project (Grant Agreement No. 101287685), funded by the SESAR 3 Joint Undertaking, and from the Spanish Ministry MICIU/AEI (Agencia Estatal de Investigación) under the Knowledge Generation Projects programme, grant PID2024-155693NB-C43 (ATHENA-AEGIS project). Funding was also received through grant FPU23/02196 funded by MICIU/AEI/10.13039/501100011033 and, as appropriate, by ``ESF Investing in your future'' or by ``European Union NextGenerationEU/PRTR''.

Views and opinions expressed in this work are those of the authors only and do not necessarily reflect those of the European Union, the SESAR 3 Joint Undertaking, any of the granting authorities, or any of the hardware providers (including IBM, IQM, Rigetti, AQT, CESGA, and AWS). The European Union, the granting authorities, nor the hardware providers can be held responsible for them.

Finally, S.H. thanks Danel Arias for his valuable insights and support regarding IBM Quantum's job scheduling and QPU architecture.

\bibliographystyle{iopart-num-titles}
\bibliography{references}      
\newpage
\appendix
\section{GHZ benchmark reports}\label{app:ghz}
\label{benchmark reports}
This appendix provides the individual GHZ benchmark reports for all evaluated quantum processors. Each report summarizes the hardware and software characteristics, experimental configuration, and GHZ benchmark results. These reports complement the main text by providing detailed information on each experimental execution.



\subsection{Processor \texttt{ibm\_basquecountry}}
\vspace{1cm}

\thispagestyle{empty}
\sffamily\small

\noindent\fbox{%
\begin{minipage}{\dimexpr\textwidth-2\fboxsep-2\fboxrule\relax}
\vspace{1pt}

\begin{center}
{\large\bfseries SPEQC / GHZ Benchmark Report}
\end{center}
\vspace{-4pt}
\hrule height 0.35pt
\vspace{3pt}

\noindent
IBM Quantum -- \texttt{ibm\_basquecountry}\\
IBM Quantum -- Basque Country\\
(Heron r2 processor, 156 qubits)

\vspace{3pt}
\hrule height 0.35pt
\vspace{3pt}

\noindent
\begin{tabular*}{\linewidth}{@{\extracolsep{\fill}} l l l l @{}}
\textbf{IR Language support:} & OpenQASM 3 & \textbf{Test date:}             & 2026-07-14 \\
\textbf{Test sponsor:}        & BIQAIN & \textbf{Hardware availability:} & 10/25 - \\
\textbf{Tested by:}           & Research team~\cite{nquire2026benchmarkingcode}& \textbf{Software version:} & 2.5 \\
\end{tabular*}

\vspace{4pt}

\noindent
\begin{minipage}[t]{0.49\linewidth}
\centering
{\normalsize\bfseries Hardware}\\[-2pt]
\dotfill
\end{minipage}\hfill
\begin{minipage}[t]{0.49\linewidth}
\centering
{\normalsize\bfseries Software}\\[-2pt]
\dotfill
\end{minipage}

\vspace{1pt}

\noindent
\begin{minipage}[t]{0.49\linewidth}
\begin{tabular}{@{}l l@{}}
\textbf{Processor name:}            & Heron r2 \\
\textbf{Backend name:}              & \texttt{ibm\_basquecountry} \\
\textbf{Location:}                  & San Sebastian, Spain \\
\textbf{Native gates:}              & X, RZ, ID, CZ, SX \\
\textbf{2Q gate (median):}          & CZ, err 0.183\%, 68.0 ns \\
\textbf{1Q gate (median):}          & SX, err 0.0361\%, 32.0 ns \\
\textbf{Readout error (median):}     & 0.647\%, 2.18 $\mu$s \\
\textbf{Topology:}                  & Heavy-hex ($\langle$deg$\rangle$=2.26) \\
\textbf{Total qubits:}              & 156 \\
\textbf{Qubits tested (GHZ size):}  & 16, 23, 24, 25, 27 \\
\textbf{Median T1:}                 & 253.1 $\mu$s \\
\textbf{Median T2:}                 & 204.6 $\mu$s \\
\end{tabular}
\end{minipage}\hfill
\begin{minipage}[t]{0.49\linewidth}
\begin{tabular}{@{}l l@{}}
\textbf{Framework:}        & Qiskit \\
\textbf{Runtime:}          & IBM Quantum Runtime \\
\textbf{Transpiler:}       & Qiskit transpiler \\
\textbf{Optimization level:} & 3 \\
\textbf{Execution mode:}   & Runtime Session (Sampler) \\
\textbf{System state:}     & Multi-user quantum backend \\
\textbf{Repository:}       & \texttt{Cross-Platform-Benchmarking*} \\
\end{tabular}
\end{minipage}

\vspace{4pt}
\hrule height 0.35pt

\begin{center}
{\normalsize\bfseries Results Table}
\end{center}
\vspace{-9pt}
\hrule height 0.35pt
\vspace{3pt}

\scriptsize
\renewcommand{\arraystretch}{1.08}
\noindent
\begin{tabular*}{\linewidth}{@{\extracolsep{\fill}} 
L{1.3cm} 
| C{0.7cm} 
C{0.95cm} 
C{1.3cm} 
| C{1.3cm} 
| C{0.75cm} 
C{1.3cm} 
C{1.2cm} 
@{}}
\textcolor{linkblue}{\underline{\bfseries Benchmark}}
& \textcolor{linkblue}{\bfseries Qubits}
& \textcolor{linkblue}{\makecell{\bfseries Depth}}
& \textcolor{linkblue}{\makecell{\bfseries 2Q gates}}
& \textcolor{linkblue}{\makecell{\bfseries GHZ\\\bfseries fidelity}}
& \textcolor{linkblue}{\bfseries Shots}
& \textcolor{linkblue}{\makecell{\bfseries Target\\\bfseries threshold}}
& \textcolor{linkblue}{\makecell{\bfseries Pass /\\\bfseries Fail}}
\\
\hline
GHZ & 16 & 27 & 15 & \textbf{0.6850} & 2952 & 0.40 & Pass \\
GHZ & 23 & 37 & 22 & \textbf{0.4729} & 2952 & 0.40 & Pass \\
GHZ & 24 & 39 & 23 & \textbf{0.4295} & 2952 & 0.40 & Pass \\
GHZ & 25 & 40 & 24 & \textbf{0.3225} & 2952 & 0.40 & \textbf{Fail} \\
GHZ & 27 & 43 & 26 & \textbf{0.2852} & 2952 & 0.40 & \textbf{Fail} \\
\hline
\multicolumn{5}{r|}{\textbf{Total circuits executed}} & \multicolumn{3}{c}{14\,760} \\
\multicolumn{5}{r|}{\textbf{Total shots executed}}    & \multicolumn{3}{c}{14\,760} \\
\hline
\multicolumn{8}{c}{\itshape Sorted by register size. Execution order (binary search): 16, 23, 27, 25, 24.} \\
\end{tabular*}

\normalsize
\vspace{4pt}

\noindent
\scriptsize
\textit{Note.} The reported GHZ fidelity corresponds to the fidelity estimator $Y$, computed from the sample mean $\hat{\mu}$ following the QuSquare benchmark suite~\cite{aguirre2025qusquare} and Eq.~\ref{eq:Y}. The estimate is obtained from a single run per register size. For each size, $l = \lceil 8\ln(4/\delta)/\varepsilon^2\rceil = 2952$ stabilizers of the GHZ state are sampled uniformly at random and each one is measured with a single shot; the 2952 measurement settings are submitted as one parametrized circuit, so each register size accounts for 2952 circuit executions and 2952 shots ($\varepsilon=\delta=0.10$). A register passes when the estimated fidelity satisfies the non-exclusion criterion of Eq.~\eqref{eq:criterion}, $Y\geq 0.5-\varepsilon=0.40$. Circuit depth and 2Q gate counts are taken from the transpiled GHZ-preparation circuits actually executed (optimization level 3, best of four GHZ layouts by CZ count), excluding the measurement-basis rotations and readout; calibration snapshot 2026-07-14 16:49:19.

\normalsize
\vspace{4pt}
\hrule height 0.35pt

\begin{center}
{\normalsize\bfseries Transpiler notes}
\end{center}
\vspace{-9pt}
\hrule height 0.35pt
\vspace{3pt}

\noindent\scriptsize\ttfamily
The Qiskit transpiler is available at \url{https://github.com/Qiskit/qiskit}
\normalfont

\vspace{4pt}
\hrule height 0.35pt

\begin{center}
{\normalsize\bfseries Submit Notes}
\end{center}
\vspace{-9pt}
\hrule height 0.35pt
\vspace{3pt}

\noindent\scriptsize\ttfamily
*These experiments were executed using the configuration files defined in \url{https://github.com/NQUIRE-Center/Cross-Platform-Benchmarking/tree/main/GHZ-State_Generation}.\\
For each register size, the GHZ-state preparation circuit was transpiled to the native gate set and connectivity of the backend, then executed for the specified number of shots.\\
The reported fidelity corresponds to the estimator value obtained for each register size, following the procedure defined in the QuSquare benchmark suite.
\normalfont

\vspace{2pt}
\end{minipage}}

\clearpage

\subsection{Processor \texttt{ibm\_pittsburgh}}
\vspace{1cm}

\thispagestyle{empty}
\sffamily\small

\noindent\fbox{%
\begin{minipage}{\dimexpr\textwidth-2\fboxsep-2\fboxrule\relax}
\vspace{1pt}

\begin{center}
{\large\bfseries SPEQC / GHZ Benchmark Report}
\end{center}
\vspace{-4pt}
\hrule height 0.35pt
\vspace{3pt}

\noindent
IBM Quantum -- \texttt{ibm\_pittsburgh}\\
IBM Quantum -- Pittsburgh\\
(Heron r3 processor, 156 qubits)

\vspace{3pt}
\hrule height 0.35pt
\vspace{3pt}

\noindent
\begin{tabular*}{\linewidth}{@{\extracolsep{\fill}} l l l l @{}}
\textbf{IR Language support:} & OpenQASM 3 & \textbf{Test date:}             & 2026-07-14 \\
\textbf{Test sponsor:}        & BIQAIN & \textbf{Hardware availability:} & 06/24 - \\
\textbf{Tested by:}           & Research team~\cite{nquire2026benchmarkingcode} & \textbf{Software version:} & 2.5 \\
\end{tabular*}

\vspace{4pt}

\noindent
\begin{minipage}[t]{0.49\linewidth}
\centering
{\normalsize\bfseries Hardware}\\[-2pt]
\dotfill
\end{minipage}\hfill
\begin{minipage}[t]{0.49\linewidth}
\centering
{\normalsize\bfseries Software}\\[-2pt]
\dotfill
\end{minipage}

\vspace{1pt}

\noindent
\begin{minipage}[t]{0.49\linewidth}
\begin{tabular}{@{}l l@{}}
\textbf{Processor name:}            & Heron r3 \\
\textbf{Backend name:}              & \texttt{ibm\_pittsburgh} \\
\textbf{Location:}                  & Poughkeepsie, United States \\
\textbf{Native gates:}              & X, RZ, ID, CZ, SX \\
\textbf{2Q gate (median):}          & CZ, err 0.157\%, 88.0 ns \\
\textbf{1Q gate (median):}          & SX, err 0.0193\%, 32.0 ns \\
\textbf{Readout error (median):}     & 0.415\%, 2.18 $\mu$s \\
\textbf{Topology:}                  & Heavy-hex ($\langle$deg$\rangle$=2.26) \\
\textbf{Total qubits:}              & 156 \\
\textbf{Qubits tested (GHZ size):}  & 16, 19, 21, 22, 23 \\
\textbf{Median T1:}                 & 284.8 $\mu$s \\
\textbf{Median T2:}                 & 311.7 $\mu$s \\
\end{tabular}
\end{minipage}\hfill
\begin{minipage}[t]{0.49\linewidth}
\begin{tabular}{@{}l l@{}}
\textbf{Framework:}        & Qiskit \\
\textbf{Runtime:}          & IBM Quantum Runtime \\
\textbf{Transpiler:}       & Qiskit transpiler \\
\textbf{Optimization level:} & 3 \\
\textbf{Execution mode:}   & Runtime Session (Sampler) \\
\textbf{System state:}     & Multi-user quantum backend \\
\textbf{Repository:}       & \texttt{Cross-Platform-Benchmarking*} \\
\end{tabular}
\end{minipage}

\vspace{4pt}
\hrule height 0.35pt

\begin{center}
{\normalsize\bfseries Results Table}
\end{center}
\vspace{-9pt}
\hrule height 0.35pt
\vspace{3pt}

\scriptsize
\renewcommand{\arraystretch}{1.08}
\noindent
\begin{tabular*}{\linewidth}{@{\extracolsep{\fill}} 
L{1.3cm} 
| C{0.7cm} 
C{0.95cm} 
C{1.3cm} 
| C{1.3cm} 
| C{0.75cm} 
C{1.3cm} 
C{1.2cm} 
@{}}
\textcolor{linkblue}{\underline{\bfseries Benchmark}}
& \textcolor{linkblue}{\bfseries Qubits}
& \textcolor{linkblue}{\makecell{\bfseries Depth}}
& \textcolor{linkblue}{\makecell{\bfseries 2Q gates}}
& \textcolor{linkblue}{\makecell{\bfseries GHZ\\\bfseries fidelity}}
& \textcolor{linkblue}{\bfseries Shots}
& \textcolor{linkblue}{\makecell{\bfseries Target\\\bfseries threshold}}
& \textcolor{linkblue}{\makecell{\bfseries Pass /\\\bfseries Fail}}
\\
\hline
GHZ & 16 & 27 & 15 & \textbf{0.5596} & 2952 & 0.40 & Pass \\
GHZ & 19 & 31 & 18 & \textbf{0.5745} & 2952 & 0.40 & Pass \\
GHZ & 21 & 34 & 20 & \textbf{0.5095} & 2952 & 0.40 & Pass \\
GHZ & 22 & 36 & 21 & \textbf{0.2954} & 2952 & 0.40 & \textbf{Fail} \\
GHZ & 23 & 37 & 22 & \textbf{0.3476} & 2952 & 0.40 & \textbf{Fail} \\
\hline
\multicolumn{5}{r|}{\textbf{Total circuits executed}} & \multicolumn{3}{c}{14\,760} \\
\multicolumn{5}{r|}{\textbf{Total shots executed}}    & \multicolumn{3}{c}{14\,760} \\
\hline
\multicolumn{8}{c}{\itshape Sorted by register size. Execution order (binary search): 16, 23, 19, 21, 22.} \\
\end{tabular*}

\normalsize
\vspace{4pt}

\noindent
\scriptsize
\textit{Note.}
The reported GHZ fidelity corresponds to the fidelity estimator $Y$, computed from the sample mean $\hat{\mu}$ following the QuSquare benchmark suite~\cite{aguirre2025qusquare} and Eq.~\ref{eq:Y}. The estimate is obtained from a single run per register size. For each size, $l = \lceil 8\ln(4/\delta)/\varepsilon^2\rceil = 2952$ stabilizers of the GHZ state are sampled uniformly at random and each one is measured with a single shot; the 2952 measurement settings are submitted as one parametrized circuit, so each register size accounts for 2952 circuit executions and 2952 shots ($\varepsilon=\delta=0.10$).A register passes when the estimated fidelity satisfies the non-exclusion criterion of Eq.~\eqref{eq:criterion}, $Y\geq 0.5-\varepsilon=0.40$. Circuit depth and 2Q gate counts are taken from the transpiled GHZ-preparation circuits actually executed (optimization level 3, best of four GHZ layouts by CZ count), excluding the measurement-basis rotations and readout; calibration snapshot 2026-07-14 15:46:51.

\normalsize
\vspace{4pt}
\hrule height 0.35pt

\begin{center}
{\normalsize\bfseries Transpiler notes}
\end{center}
\vspace{-9pt}
\hrule height 0.35pt
\vspace{3pt}

\noindent\scriptsize\ttfamily
The Qiskit transpiler is available at \url{https://github.com/Qiskit/qiskit}
\normalfont

\vspace{4pt}
\hrule height 0.35pt

\begin{center}
{\normalsize\bfseries Submit Notes}
\end{center}
\vspace{-9pt}
\hrule height 0.35pt
\vspace{3pt}

\noindent\scriptsize\ttfamily
*These experiments were executed using the configuration files defined in \url{https://github.com/NQUIRE-Center/Cross-Platform-Benchmarking/tree/main/GHZ-State_Generation}.\\
For each register size, the GHZ-state preparation circuit was transpiled to the native gate set and connectivity of the backend, then executed for the specified number of shots.\\
The reported fidelity corresponds to the estimator value obtained for each register size, following the procedure defined in the QuSquare benchmark suite.
\normalfont

\vspace{2pt}
\end{minipage}}

\clearpage

\subsection{Processor \texttt{ibm\_miami}}
\vspace{1cm}

\thispagestyle{empty}
\sffamily\small

\noindent\fbox{%
\begin{minipage}{\dimexpr\textwidth-2\fboxsep-2\fboxrule\relax}
\vspace{1pt}

\begin{center}
{\large\bfseries SPEQC / GHZ Benchmark Report}
\end{center}
\vspace{-4pt}
\hrule height 0.35pt
\vspace{3pt}

\noindent
IBM Quantum -- \texttt{ibm\_miami}\\
IBM Quantum -- Miami\\
(Nighthawk r1 processor, 120 qubits)

\vspace{3pt}
\hrule height 0.35pt
\vspace{3pt}

\noindent
\begin{tabular*}{\linewidth}{@{\extracolsep{\fill}} l l l l @{}}
\textbf{IR Language support:} & OpenQASM 3 & \textbf{Test date:}             & 2026-07-14 \\
\textbf{Test sponsor:}        & BIQAIN & \textbf{Hardware availability:} & 01/26 - \\
\textbf{Tested by:}           & Research team~\cite{nquire2026benchmarkingcode} & \textbf{Software version:} & 2.5 \\
\end{tabular*}

\vspace{4pt}

\noindent
\begin{minipage}[t]{0.49\linewidth}
\centering
{\normalsize\bfseries Hardware}\\[-2pt]
\dotfill
\end{minipage}\hfill
\begin{minipage}[t]{0.49\linewidth}
\centering
{\normalsize\bfseries Software}\\[-2pt]
\dotfill
\end{minipage}

\vspace{1pt}

\noindent
\begin{minipage}[t]{0.49\linewidth}
\begin{tabular}{@{}l l@{}}
\textbf{Processor name:}            & Nighthawk r1 \\
\textbf{Backend name:}              & \texttt{ibm\_miami} \\
\textbf{Location:}                  & Poughkeepsie, United States \\
\textbf{Native gates:}              & X, RZ, ID, CZ, SX \\
\textbf{2Q gate (median):}          & CZ, err 0.319\%, 138.0 ns \\
\textbf{1Q gate (median):}          & SX, err 0.0240\%, 32.0 ns \\
\textbf{Readout error (median):}     & 1.99\%, 2.40 $\mu$s \\
\textbf{Topology:}                  & Square grid ($\langle$deg$\rangle$=3.63) \\
\textbf{Total qubits:}              & 120 \\
\textbf{Qubits tested (GHZ size):}  & 16, 17, 18, 19, 23 \\
\textbf{Median T1:}                 & 304.4 $\mu$s \\
\textbf{Median T2:}                 & 205.2 $\mu$s \\
\end{tabular}
\end{minipage}\hfill
\begin{minipage}[t]{0.49\linewidth}
\begin{tabular}{@{}l l@{}}
\textbf{Framework:}        & Qiskit \\
\textbf{Runtime:}          & IBM Quantum Runtime \\
\textbf{Transpiler:}       & Qiskit transpiler \\
\textbf{Optimization level:} & 3 \\
\textbf{Execution mode:}   & Runtime Session (Sampler) \\
\textbf{System state:}     & Multi-user quantum backend \\
\textbf{Repository:}       & \texttt{Cross-Platform-Benchmarking*} \\
\end{tabular}
\end{minipage}

\vspace{4pt}
\hrule height 0.35pt

\begin{center}
{\normalsize\bfseries Results Table}
\end{center}
\vspace{-9pt}
\hrule height 0.35pt
\vspace{3pt}

\scriptsize
\renewcommand{\arraystretch}{1.08}
\noindent
\begin{tabular*}{\linewidth}{@{\extracolsep{\fill}} 
L{1.3cm} 
| C{0.7cm} 
C{0.95cm} 
C{1.3cm} 
| C{1.3cm} 
| C{0.75cm} 
C{1.3cm} 
C{1.2cm} 
@{}}
\textcolor{linkblue}{\underline{\bfseries Benchmark}}
& \textcolor{linkblue}{\bfseries Qubits}
& \textcolor{linkblue}{\makecell{\bfseries Depth}}
& \textcolor{linkblue}{\makecell{\bfseries 2Q gates}}
& \textcolor{linkblue}{\makecell{\bfseries GHZ\\\bfseries fidelity}}
& \textcolor{linkblue}{\bfseries Shots}
& \textcolor{linkblue}{\makecell{\bfseries Target\\\bfseries threshold}}
& \textcolor{linkblue}{\makecell{\bfseries Pass /\\\bfseries Fail}}
\\
\hline
GHZ & 16 & 27 & 15 & \textbf{0.4715} & 2952 & 0.40 & Pass \\
GHZ & 17 & 28 & 16 & \textbf{0.5169} & 2952 & 0.40 & Pass \\
GHZ & 18 & 30 & 17 & \textbf{0.4228} & 2952 & 0.40 & Pass \\
GHZ & 19 & 31 & 18 & \textbf{0.3787} & 2952 & 0.40 & \textbf{Fail} \\
GHZ & 23 & 37 & 22 & \textbf{0.1687} & 2952 & 0.40 & \textbf{Fail} \\
\hline
\multicolumn{5}{r|}{\textbf{Total circuits executed}} & \multicolumn{3}{c}{14\,760} \\
\multicolumn{5}{r|}{\textbf{Total shots executed}}    & \multicolumn{3}{c}{14\,760} \\
\hline
\multicolumn{8}{c}{\itshape Sorted by register size. Execution order (binary search): 16, 23, 19, 17, 18.} \\
\end{tabular*}

\normalsize
\vspace{4pt}

\noindent
\scriptsize
\textit{Note.} The reported GHZ fidelity corresponds to the fidelity estimator $Y$, computed from the sample mean $\hat{\mu}$ following the QuSquare benchmark suite~\cite{aguirre2025qusquare} and Eq.~\ref{eq:Y}. The estimate is obtained from a single run per register size. For each size, $l = \lceil 8\ln(4/\delta)/\varepsilon^2\rceil = 2952$ stabilizers of the GHZ state are sampled uniformly at random and each one is measured with a single shot; the 2952 measurement settings are submitted as one parametrized circuit, so each register size accounts for 2952 circuit executions and 2952 shots ($\varepsilon=\delta=0.10$). A register passes when the estimated fidelity satisfies the non-exclusion criterion of Eq.~\eqref{eq:criterion}, $Y\geq 0.5-\varepsilon=0.40$. Circuit depth and 2Q gate counts are taken from the transpiled GHZ-preparation circuits actually executed (optimization level 3, best of four GHZ layouts by CZ count), excluding the measurement-basis rotations and readout; calibration snapshot 2026-07-14 16:24:10.

\normalsize
\vspace{4pt}
\hrule height 0.35pt

\begin{center}
{\normalsize\bfseries Transpiler notes}
\end{center}
\vspace{-9pt}
\hrule height 0.35pt
\vspace{3pt}

\noindent\scriptsize\ttfamily
The Qiskit transpiler is available at \url{https://github.com/Qiskit/qiskit}
\normalfont

\vspace{4pt}
\hrule height 0.35pt

\begin{center}
{\normalsize\bfseries Submit Notes}
\end{center}
\vspace{-9pt}
\hrule height 0.35pt
\vspace{3pt}

\noindent\scriptsize\ttfamily
*These experiments were executed using the configuration files defined in \url{https://github.com/NQUIRE-Center/Cross-Platform-Benchmarking/tree/main/GHZ-State_Generation}.\\
For each register size, the GHZ-state preparation circuit was transpiled to the native gate set and connectivity of the backend, then executed for the specified number of shots.\\
The reported fidelity corresponds to the estimator value obtained for each register size, following the procedure defined in the QuSquare benchmark suite.
\normalfont

\vspace{2pt}
\end{minipage}}

\clearpage

\subsection{Processor \texttt{ibm\_phoenix}}
\vspace{1cm}

\thispagestyle{empty}
\sffamily\small

\noindent\fbox{%
\begin{minipage}{\dimexpr\textwidth-2\fboxsep-2\fboxrule\relax}
\vspace{1pt}

\begin{center}
{\large\bfseries SPEQC / GHZ Benchmark Report}
\end{center}
\vspace{-4pt}
\hrule height 0.35pt
\vspace{3pt}

\noindent
IBM Quantum -- \texttt{ibm\_phoenix}\\
IBM Quantum -- Phoenix\\
(Nighthawk r2 processor, 120 qubits)

\vspace{3pt}
\hrule height 0.35pt
\vspace{3pt}

\noindent
\begin{tabular*}{\linewidth}{@{\extracolsep{\fill}} l l l l @{}}
\textbf{IR Language support:} & OpenQASM 3 & \textbf{Test date:}             & 2026-09-03 \\
\textbf{Test sponsor:}        & BIQAIN & \textbf{Hardware availability:} & 09/26 -  \\
\textbf{Tested by:}           & Research team~\cite{nquire2026benchmarkingcode} & \textbf{Software version:} & 2.5 \\
\end{tabular*}

\vspace{4pt}

\noindent
\begin{minipage}[t]{0.49\linewidth}
\centering
{\normalsize\bfseries Hardware}\\[-2pt]
\dotfill
\end{minipage}\hfill
\begin{minipage}[t]{0.49\linewidth}
\centering
{\normalsize\bfseries Software}\\[-2pt]
\dotfill
\end{minipage}

\vspace{1pt}

\noindent
\begin{minipage}[t]{0.49\linewidth}
\begin{tabular}{@{}l l@{}}
\textbf{Processor name:}            & Nighthawk r2 \\
\textbf{Backend name:}              & \texttt{ibm\_phoenix} \\
\textbf{Location:}                  & Poughkeepsie, United States \\
\textbf{Native gates:}              & X, RZ, ID, CZ, SX \\
\textbf{2Q gate (median):}          & CZ, err 0.203\%, 68.0 ns \\
\textbf{1Q gate (median):}          & SX, err 0.0217\%, 40.0 ns \\
\textbf{Readout error (median):}     & 0.732\%, 1.94 $\mu$s \\
\textbf{Topology:}                  & Square grid ($\langle$deg$\rangle$=3.63) \\
\textbf{Total qubits:}              & 120 \\
\textbf{Qubits tested (GHZ size):}  & 16, 23, 27, 28, 29 \\
\textbf{Median T1:}                 & 175.6 $\mu$s \\
\textbf{Median T2:}                 & 139.9 $\mu$s \\
\end{tabular}
\end{minipage}\hfill
\begin{minipage}[t]{0.49\linewidth}
\begin{tabular}{@{}l l@{}}
\textbf{Framework:}        & Qiskit \\
\textbf{Runtime:}          & IBM Quantum Runtime \\
\textbf{Transpiler:}       & Qiskit transpiler \\
\textbf{Optimization level:} & 3 \\
\textbf{Execution mode:}   & Runtime Session (Sampler) \\
\textbf{System state:}     & Multi-user quantum backend \\
\textbf{Repository:}       & \texttt{Cross-Platform-Benchmarking*} \\
\end{tabular}
\end{minipage}

\vspace{4pt}
\hrule height 0.35pt

\begin{center}
{\normalsize\bfseries Results Table}
\end{center}
\vspace{-9pt}
\hrule height 0.35pt
\vspace{3pt}

\scriptsize
\renewcommand{\arraystretch}{1.08}
\noindent
\begin{tabular*}{\linewidth}{@{\extracolsep{\fill}} 
L{1.3cm} 
| C{0.7cm} 
C{0.95cm} 
C{1.3cm} 
| C{1.3cm} 
| C{0.75cm} 
C{1.3cm} 
C{1.2cm} 
@{}}
\textcolor{linkblue}{\underline{\bfseries Benchmark}}
& \textcolor{linkblue}{\bfseries Qubits}
& \textcolor{linkblue}{\makecell{\bfseries Depth}}
& \textcolor{linkblue}{\makecell{\bfseries 2Q gates}}
& \textcolor{linkblue}{\makecell{\bfseries GHZ\\\bfseries fidelity}}
& \textcolor{linkblue}{\bfseries Shots}
& \textcolor{linkblue}{\makecell{\bfseries Target\\\bfseries threshold}}
& \textcolor{linkblue}{\makecell{\bfseries Pass /\\\bfseries Fail}}
\\
\hline
GHZ & 16 & 27 & 15 & \textbf{0.7398} & 2952 & 0.40 & Pass \\
GHZ & 23 & 37 & 22 & \textbf{0.5081} & 2952 & 0.40 & Pass \\
GHZ & 27 & 43 & 26 & \textbf{0.4377} & 2952 & 0.40 & Pass \\
GHZ & 28 & 45 & 27 & \textbf{0.3150} & 2952 & 0.40 & \textbf{Fail} \\
GHZ & 29 & 46 & 28 & \textbf{0.3449} & 2952 & 0.40 & \textbf{Fail} \\
\hline
\multicolumn{5}{r|}{\textbf{Total circuits executed}} & \multicolumn{3}{c}{14\,760} \\
\multicolumn{5}{r|}{\textbf{Total shots executed}}    & \multicolumn{3}{c}{14\,760} \\
\hline
\multicolumn{8}{c}{\itshape Sorted by register size. Execution order (binary search): 16, 23, 27, 29, 28.} \\
\end{tabular*}

\normalsize
\vspace{4pt}

\noindent
\scriptsize
\textit{Note.} 
The reported GHZ fidelity corresponds to the fidelity estimator $Y$, computed from the sample mean $\hat{\mu}$ following the QuSquare benchmark suite~\cite{aguirre2025qusquare} and Eq.~\ref{eq:Y}. The estimate is obtained from a single run per register size. For each size, $l = \lceil 8\ln(4/\delta)/\varepsilon^2\rceil = 2952$ stabilizers of the GHZ state are sampled uniformly at random and each one is measured with a single shot; the 2952 measurement settings are submitted as one parametrized circuit, so each register size accounts for 2952 circuit executions and 2952 shots ($\varepsilon=\delta=0.10$). A register passes when the estimated fidelity satisfies the non-exclusion criterion of Eq.~\eqref{eq:criterion}, $Y\geq 0.5-\varepsilon=0.40$. Circuit depth and 2Q gate counts are taken from the transpiled GHZ-preparation circuits actually executed (optimization level 3, best of four GHZ layouts by CZ count), excluding the measurement-basis rotations and readout; calibration snapshot 2026-09-03 09:37:40. Median T1 and T2 exclude qubit 17, which reported no coherence data in this calibration.

\normalsize
\vspace{4pt}
\hrule height 0.35pt

\begin{center}
{\normalsize\bfseries Transpiler notes}
\end{center}
\vspace{-9pt}
\hrule height 0.35pt
\vspace{3pt}

\noindent\scriptsize\ttfamily
The Qiskit transpiler is available at \url{https://github.com/Qiskit/qiskit}
\normalfont

\vspace{4pt}
\hrule height 0.35pt

\begin{center}
{\normalsize\bfseries Submit Notes}
\end{center}
\vspace{-9pt}
\hrule height 0.35pt
\vspace{3pt}

\noindent\scriptsize\ttfamily
*These experiments were executed using the configuration files defined in \url{https://github.com/NQUIRE-Center/Cross-Platform-Benchmarking/tree/main/GHZ-State_Generation}.\\
For each register size, the GHZ-state preparation circuit was transpiled to the native gate set and connectivity of the backend, then executed for the specified number of shots.\\
The reported fidelity corresponds to the estimator value obtained for each register size, following the procedure defined in the QuSquare benchmark suite.
\normalfont

\vspace{2pt}
\end{minipage}}

\clearpage

\subsection{Processor \texttt{iqm/Garnet}}       
\vspace{1cm}

\thispagestyle{empty}
\sffamily\small

\noindent\fbox{%
\begin{minipage}{\dimexpr\textwidth-2\fboxsep-2\fboxrule\relax}
\vspace{1pt}

\begin{center}
{\large\bfseries SPEQC / GHZ Benchmark Report}
\end{center}
\vspace{-4pt}
\hrule height 0.35pt
\vspace{3pt}

\noindent
AWS Braket -- \texttt{qpu/iqm/Garnet}\\
IQM\\
(Superconducting QPU, 20 qubits)

\vspace{3pt}
\hrule height 0.35pt
\vspace{3pt}

\noindent
\begin{tabular*}{\linewidth}{@{\extracolsep{\fill}} l l l l @{}}
\textbf{IR Language support:} & OpenQASM 3 & \textbf{Test date:}             & 2026-07-14 \\
\textbf{Test sponsor:}        & AWS & \textbf{Hardware availability:} & 05/24 -  \\
\textbf{Tested by:}           & Research team~\cite{nquire2026benchmarkingcode} & \textbf{Software version:} & 1.121.0 \\
\end{tabular*}

\vspace{4pt}

\noindent
\begin{minipage}[t]{0.49\linewidth}
\centering
{\normalsize\bfseries Hardware}\\[-2pt]
\dotfill
\end{minipage}\hfill
\begin{minipage}[t]{0.49\linewidth}
\centering
{\normalsize\bfseries Software}\\[-2pt]
\dotfill
\end{minipage}

\vspace{1pt}

\noindent
\begin{minipage}[t]{0.49\linewidth}
\begin{tabular}{@{}l l@{}}
\textbf{Processor name:}            & IQM Garnet \\
\textbf{Backend name:}              & \texttt{iqm/Garnet} \\
\textbf{Location:}                  & Espoo, Finland \\
\textbf{Native gates:}              & CZ, PRX, CC\_PRX \\
\textbf{2Q gate (median):}          & CZ, err 0.670\% \\
\textbf{1Q gate (median):}          & PRX, err 0.086\% \\
\textbf{Topology:}                  & Sparse graph ($\langle$deg$\rangle$=3.00) \\
\textbf{Total qubits:}              & 20 \\
\textbf{Qubits tested (GHZ size):}  & 8, 12, 13, 14, 16 \\
\textbf{Median T1:}                 & 32.1 $\mu$s \\
\textbf{Median T2:}                 & 9.0 $\mu$s \\
\end{tabular}
\end{minipage}\hfill
\begin{minipage}[t]{0.49\linewidth}
\begin{tabular}{@{}l l@{}}
\textbf{Framework:}        & Amazon Braket SDK \\
\textbf{Runtime:}          & AWS Braket  \\
\textbf{Transpiler:}       & Default provider compilation \\
\textbf{Optimization level:} & Not specified \\
\textbf{Execution mode:}   & On-demand (ProgramSet) \\
\textbf{System state:}     & Multi-user quantum backend \\
\textbf{Repository:}       & \texttt{Cross-Platform-Benchmarking*} \\
\end{tabular}
\end{minipage}

\vspace{4pt}
\hrule height 0.35pt

\begin{center}
{\normalsize\bfseries Results Table}
\end{center}
\vspace{-9pt}
\hrule height 0.35pt
\vspace{3pt}

\noindent\scriptsize
\textbf{A single run was executed per register size (binary search). Depth and 2Q gates are LOGICAL values (pre-transpilation circuit), not the exact native-gate values executed -- see Note.}
\normalsize

\vspace{3pt}

\scriptsize
\renewcommand{\arraystretch}{1.08}
\noindent
\begin{tabular*}{\linewidth}{@{\extracolsep{\fill}} 
L{1.3cm} 
| C{0.8cm} 
C{1.3cm} 
C{1.5cm} 
| C{1.4cm} 
| C{1.1cm} 
C{1.6cm} 
C{1.0cm} 
@{}}
\textcolor{linkblue}{\underline{\bfseries Benchmark}}
& \textcolor{linkblue}{\bfseries Qubits}
& \textcolor{linkblue}{\makecell{\bfseries Depth$^\dagger$}}
& \textcolor{linkblue}{\makecell{\bfseries 2Q gates$^\dagger$}}
& \textcolor{linkblue}{\makecell{\bfseries GHZ\\\bfseries fidelity}}
& \textcolor{linkblue}{\makecell{\bfseries Shots}}
& \textcolor{linkblue}{\makecell{\bfseries Target\\\bfseries threshold}}
& \textcolor{linkblue}{\makecell{\bfseries Pass /\\\bfseries Fail}}
\\
\hline
GHZ & 8 & 6 & 7 & \textbf{0.7038} & 2952 & 0.40 & Pass \\
GHZ & 12 & 8 & 11 & \textbf{0.4926} & 2952 & 0.40 & Pass \\
GHZ & 13 & 9 & 12 & \textbf{0.4371} & 2952 & 0.40 & Pass \\
GHZ & 14 & 9 & 13 & \textbf{0.3787} & 2952 & 0.40 & \textbf{Fail} \\
GHZ & 16 & 10 & 15 & \textbf{0.3659} & 2952 & 0.40 & \textbf{Fail} \\
\hline
\multicolumn{4}{r|}{\textbf{Total circuits executed}} & \multicolumn{4}{c}{5} \\
\multicolumn{4}{r|}{\textbf{Total shots executed}}    & \multicolumn{4}{c}{14\,760} \\
\hline
\multicolumn{8}{c}{\itshape Sorted by register size. Execution order (binary search): 16, 8, 12, 14, 13.} \\
\end{tabular*}

\normalsize
\vspace{4pt}

\noindent
\scriptsize
\textit{Note.} 
The reported GHZ fidelity corresponds to the fidelity estimator $Y$, computed from the sample mean $\hat{\mu}$ following the QuSquare benchmark suite~\cite{aguirre2025qusquare} and Eq.~\ref{eq:Y}. For each register size, $l$ non-identity GHZ stabilizers are sampled uniformly at random, each measured $m_i$ times, so shots $=l\times m_i$. Here $l = \lceil 8\ln(4/\delta)/\varepsilon^2\rceil$ with $\varepsilon=\delta=0.10$, giving $l=2952$. Here $m_i=1$, so shots $=2952$ for every row. A register passes when the estimated fidelity satisfies the non-exclusion criterion of Eq.~\eqref{eq:criterion}, $Y\geq 0.5-\varepsilon=0.40$. $^\dagger$Depth/2Q gates are computed on the \textbf{logical} (pre-transpilation, H+CNOT) circuit saved per run -- not the real hardware-native circuit, which was not recorded per-shot by the execution harness (field \texttt{circuit} is hardcoded to \texttt{"Not available in program\_set"} in \texttt{aws\_benchmark.py}).  Hardware figures are the medians of the backend calibration snapshot dated 2026-07-14 (device location: Espoo, Finland), taken shortly before the run (2026-07-14).

\normalsize
\vspace{4pt}
\hrule height 0.35pt

\begin{center}
{\normalsize\bfseries Submit Notes}
\end{center}
\vspace{-9pt}
\hrule height 0.35pt
\vspace{3pt}

\noindent\scriptsize\ttfamily
*These experiments were executed using the configuration files defined in \url{https://github.com/NQUIRE-Center/Cross-Platform-Benchmarking/tree/main/GHZ-State_Generation}.\\
The GHZ experiments were executed through Amazon Braket using the Amazon Braket SDK's \texttt{ProgramSet}/\texttt{CircuitBinding} API.\\
For each register size, a single run was performed. The GHZ-state preparation circuit (linear H+CNOT ladder) was submitted to the backend using the default provider compilation to its native gate set.\\
\normalfont

\vspace{2pt}
\end{minipage}}

\clearpage

\subsection{Processor \texttt{iqm/Emerald}}   
\vspace{1cm}

\thispagestyle{empty}
\sffamily\small

\noindent\fbox{%
\begin{minipage}{\dimexpr\textwidth-2\fboxsep-2\fboxrule\relax}
\vspace{1pt}

\begin{center}
{\large\bfseries SPEQC / GHZ Benchmark Report}
\end{center}
\vspace{-4pt}
\hrule height 0.35pt
\vspace{3pt}

\noindent
AWS Braket -- \texttt{qpu/iqm/Emerald}\\
IQM\\
(Superconducting QPU, 54 qubits)

\vspace{3pt}
\hrule height 0.35pt
\vspace{3pt}

\noindent
\begin{tabular*}{\linewidth}{@{\extracolsep{\fill}} l l l l @{}}
\textbf{IR Language support:} & OpenQASM 3 & \textbf{Test date:}             & 2026-07-14 \\
\textbf{Test sponsor:}        & AWS & \textbf{Hardware availability:} & 07/25 - \\
\textbf{Tested by:}           & Research team~\cite{nquire2026benchmarkingcode} & \textbf{Software version:} & 1.121.0 \\
\end{tabular*}

\vspace{4pt}

\noindent
\begin{minipage}[t]{0.49\linewidth}
\centering
{\normalsize\bfseries Hardware}\\[-2pt]
\dotfill
\end{minipage}\hfill
\begin{minipage}[t]{0.49\linewidth}
\centering
{\normalsize\bfseries Software}\\[-2pt]
\dotfill
\end{minipage}

\vspace{1pt}

\noindent
\begin{minipage}[t]{0.49\linewidth}
\begin{tabular}{@{}l l@{}}
\textbf{Processor name:}            & IQM Emerald \\
\textbf{Backend name:}              & \texttt{iqm/Emerald} \\
\textbf{Location:}                  & Munich, Germany \\
\textbf{Native gates:}              & CZ, PRX, CC\_PRX \\
\textbf{2Q gate (median):}          & CZ, err 0.418\% \\
\textbf{1Q gate (median):}          & PRX, err 0.034\% \\
\textbf{Topology:}                  & Sparse graph ($\langle$deg$\rangle$=3.15) \\
\textbf{Total qubits:}              & 54 \\
\textbf{Qubits tested (GHZ size):}  & 16, 19, 20, 21, 23 \\
\textbf{Median T1:}                 & 52.3 $\mu$s \\
\textbf{Median T2:}                 & 16.0 $\mu$s \\
\end{tabular}
\end{minipage}\hfill
\begin{minipage}[t]{0.49\linewidth}
\begin{tabular}{@{}l l@{}}
\textbf{Framework:}        & Amazon Braket SDK \\
\textbf{Runtime:}          & AWS Braket \\
\textbf{Transpiler:}       & Default provider compilation \\
\textbf{Optimization level:} & Not specified \\
\textbf{Execution mode:}   & On-demand (ProgramSet) \\
\textbf{System state:}     & Multi-user quantum backend \\
\textbf{Repository:}       & \texttt{Cross-Platform-Benchmarking*} \\
\end{tabular}
\end{minipage}

\vspace{4pt}
\hrule height 0.35pt

\begin{center}
{\normalsize\bfseries Results Table}
\end{center}
\vspace{-9pt}
\hrule height 0.35pt
\vspace{3pt}

\noindent\scriptsize
\textbf{A single run was executed per register size (binary search). Depth and 2Q gates are LOGICAL values (pre-transpilation circuit), not the exact native-gate values executed -- see Note.}
\normalsize

\vspace{3pt}

\scriptsize
\renewcommand{\arraystretch}{1.08}
\noindent
\begin{tabular*}{\linewidth}{@{\extracolsep{\fill}} 
L{1.3cm} 
| C{0.8cm} 
C{1.3cm} 
C{1.5cm} 
| C{1.4cm} 
| C{1.1cm} 
C{1.6cm} 
C{1.0cm} 
@{}}
\textcolor{linkblue}{\underline{\bfseries Benchmark}}
& \textcolor{linkblue}{\bfseries Qubits}
& \textcolor{linkblue}{\makecell{\bfseries Depth$^\dagger$}}
& \textcolor{linkblue}{\makecell{\bfseries 2Q gates$^\dagger$}}
& \textcolor{linkblue}{\makecell{\bfseries GHZ\\\bfseries fidelity}}
& \textcolor{linkblue}{\makecell{\bfseries Shots}}
& \textcolor{linkblue}{\makecell{\bfseries Target\\\bfseries threshold}}
& \textcolor{linkblue}{\makecell{\bfseries Pass /\\\bfseries Fail}}
\\
\hline
GHZ & 16 & 10 & 15 & \textbf{0.5603} & 2952 & 0.40 & Pass \\
GHZ & 19 & 12 & 18 & \textbf{0.4241} & 2952 & 0.40 & Pass \\
GHZ & 20 & 12 & 19 & \textbf{0.4207} & 2952 & 0.40 & Pass \\
GHZ & 21 & 13 & 20 & \textbf{0.3679} & 2952 & 0.40 & \textbf{Fail} \\
GHZ & 23 & 14 & 22 & \textbf{0.3469} & 2952 & 0.40 & \textbf{Fail} \\
\hline
\multicolumn{4}{r|}{\textbf{Total circuits executed}} & \multicolumn{4}{c}{5} \\
\multicolumn{4}{r|}{\textbf{Total shots executed}}    & \multicolumn{4}{c}{14\,760} \\
\hline
\multicolumn{8}{c}{\itshape Sorted by register size. Execution order (binary search): 16, 23, 19, 21, 20.} \\
\end{tabular*}

\normalsize
\vspace{4pt}

\noindent
\scriptsize
\textit{Note.} 
The reported GHZ fidelity corresponds to the fidelity estimator $Y$, computed from the sample mean $\hat{\mu}$ following the QuSquare benchmark suite~\cite{aguirre2025qusquare} and Eq.~\ref{eq:Y}. For each register size, $l$ non-identity GHZ stabilizers are sampled uniformly at random, each measured $m_i$ times, so shots $=l\times m_i$. Here $l = \lceil 8\ln(4/\delta)/\varepsilon^2\rceil$ with $\varepsilon=\delta=0.10$, giving $l=2952$. Here $m_i=1$, so shots $=2952$ for every row. A register passes when the estimated fidelity satisfies the non-exclusion criterion of Eq.~\eqref{eq:criterion}, $Y\geq 0.5-\varepsilon=0.40$. $^\dagger$Depth/2Q gates are computed on the \textbf{logical} (pre-transpilation, H+CNOT) circuit saved per run -- not the real hardware-native circuit, which was not recorded per-shot by the execution harness (field \texttt{circuit} is hardcoded to \texttt{"Not available in program\_set"} in \texttt{aws\_benchmark.py}). For reference, the actual native-gate (post-transpilation) circuit at $N=16$ was recovered from a separately-saved template (not from this per-shot metadata): depth=21, 2Q gates=19 (vs.\ logical 10/15), fidelity 0.5603 -- showing real routing/compilation overhead at this backend's connectivity. Hardware figures are the medians of the backend calibration snapshot dated 2026-07-14 (device location: Munich, Germany), taken shortly before the run (2026-07-14).

\normalsize
\vspace{4pt}
\hrule height 0.35pt

\begin{center}
{\normalsize\bfseries Submit Notes}
\end{center}
\vspace{-9pt}
\hrule height 0.35pt
\vspace{3pt}

\noindent\scriptsize\ttfamily
*These experiments were executed using the configuration files defined in \url{https://github.com/NQUIRE-Center/Cross-Platform-Benchmarking/tree/main/GHZ-State_Generation}.\\
The GHZ experiments were executed through Amazon Braket using the Amazon Braket SDK's \texttt{ProgramSet}/\texttt{CircuitBinding} API.\\
For each register size, a single run was performed. The GHZ-state preparation circuit (linear H+CNOT ladder) was submitted to the backend using the default provider compilation to its native gate set.\\
\normalfont

\vspace{2pt}
\end{minipage}}

\clearpage

\subsection{Processor \texttt{rigetti/Cepheus-1}}
\vspace{1cm}

\thispagestyle{empty}
\sffamily\small

\noindent\fbox{%
\begin{minipage}{\dimexpr\textwidth-2\fboxsep-2\fboxrule\relax}
\vspace{1pt}

\begin{center}
{\large\bfseries SPEQC / GHZ Benchmark Report}
\end{center}
\vspace{-4pt}
\hrule height 0.35pt
\vspace{3pt}

\noindent
AWS Braket -- \texttt{qpu/rigetti/Cepheus-1-108Q}\\
Rigetti\\
(Superconducting QPU, 107 qubits)

\vspace{3pt}
\hrule height 0.35pt
\vspace{3pt}

\noindent
\begin{tabular*}{\linewidth}{@{\extracolsep{\fill}} l l l l @{}}
\textbf{IR Language support:} & OpenQASM 3 & \textbf{Test date:}             & 2026-07-11 \\
\textbf{Test sponsor:}        & AWS & \textbf{Hardware availability:} & 05/26 - \\
\textbf{Tested by:}           & Research team~\cite{nquire2026benchmarkingcode} & \textbf{Software version:} & 1.121.0 \\
\end{tabular*}

\vspace{4pt}

\noindent
\begin{minipage}[t]{0.49\linewidth}
\centering
{\normalsize\bfseries Hardware}\\[-2pt]
\dotfill
\end{minipage}\hfill
\begin{minipage}[t]{0.49\linewidth}
\centering
{\normalsize\bfseries Software}\\[-2pt]
\dotfill
\end{minipage}

\vspace{1pt}

\noindent
\begin{minipage}[t]{0.49\linewidth}
\begin{tabular}{@{}l l@{}}
\textbf{Processor name:}            & Rigetti Cepheus-1 \\
\textbf{Backend name:}              & \texttt{rigetti/Cepheus-1-108Q} \\
\textbf{Location:}                  & California, USA \\
\textbf{Native gates:}              & RX, RZ, CZ \\
\textbf{2Q gate (median):}          & CZ, err 1.312\% \\
\textbf{1Q gate (median):}          & RX, err 0.263\% \\
\textbf{Topology:}                  & Sparse graph ($\langle$deg$\rangle$=3.61) \\
\textbf{Total qubits:}              & 107 \\
\textbf{Qubits tested (GHZ size):}  & 4, 5, 6, 8, 16 \\
\textbf{Median T1:}                 & 24.8 $\mu$s \\
\textbf{Median T2:}                 & 9.3 $\mu$s \\
\end{tabular}
\end{minipage}\hfill
\begin{minipage}[t]{0.49\linewidth}
\begin{tabular}{@{}l l@{}}
\textbf{Framework:}        & Amazon Braket SDK \\
\textbf{Runtime:}          & AWS Braket  \\
\textbf{Transpiler:}       & Default provider compilation\\
\textbf{Optimization level:}  & Not specified \\
\textbf{Execution mode:}   & On-demand (ProgramSet) \\
\textbf{System state:}     & Multi-user quantum backend \\
\textbf{Repository:}       & \texttt{Cross-Platform-Benchmarking*} \\
\end{tabular}
\end{minipage}

\vspace{4pt}
\hrule height 0.35pt

\begin{center}
{\normalsize\bfseries Results Table}
\end{center}
\vspace{-9pt}
\hrule height 0.35pt
\vspace{3pt}

\noindent\scriptsize
\textbf{A single run was executed per register size (binary search). Depth and 2Q gates are LOGICAL values (pre-transpilation circuit), not the exact native-gate values executed -- see Note.}
\normalsize

\vspace{3pt}

\scriptsize
\renewcommand{\arraystretch}{1.08}
\noindent
\begin{tabular*}{\linewidth}{@{\extracolsep{\fill}} 
L{1.3cm} 
| C{0.8cm} 
C{1.3cm} 
C{1.5cm} 
| C{1.4cm} 
| C{1.1cm} 
C{1.6cm} 
C{1.0cm} 
@{}}
\textcolor{linkblue}{\underline{\bfseries Benchmark}}
& \textcolor{linkblue}{\bfseries Qubits}
& \textcolor{linkblue}{\makecell{\bfseries Depth$^\dagger$}}
& \textcolor{linkblue}{\makecell{\bfseries 2Q gates$^\dagger$}}
& \textcolor{linkblue}{\makecell{\bfseries GHZ\\\bfseries fidelity}}
& \textcolor{linkblue}{\makecell{\bfseries Shots}}
& \textcolor{linkblue}{\makecell{\bfseries Target\\\bfseries threshold}}
& \textcolor{linkblue}{\makecell{\bfseries Pass /\\\bfseries Fail}}
\\
\hline
GHZ & 4 & 4 & 3 & \textbf{0.4906} & 29520 & 0.40 & Pass \\
GHZ & 5 & 5 & 4 & \textbf{0.2229} & 29520 & 0.40 & \textbf{Fail} \\
GHZ & 6 & 5 & 5 & \textbf{0.3311} & 29520 & 0.40 & \textbf{Fail} \\
GHZ & 8 & 6 & 7 & \textbf{0.0626} & 29520 & 0.40 & \textbf{Fail} \\
GHZ & 16 & 10 & 15 & \textbf{0.0210} & 29520 & 0.40 & \textbf{Fail} \\
\hline
\multicolumn{4}{r|}{\textbf{Total circuits executed}} & \multicolumn{4}{c}{5} \\
\multicolumn{4}{r|}{\textbf{Total shots executed}}    & \multicolumn{4}{c}{147\,600} \\
\hline
\multicolumn{8}{c}{\itshape Sorted by register size. Execution order (binary search): 16, 8, 6, 5, 4.} \\
\end{tabular*}

\normalsize
\vspace{4pt}

\noindent
\scriptsize
\textit{Note.} The reported GHZ fidelity corresponds to the fidelity estimator $Y$, computed from the sample mean $\hat{\mu}$ following the stabilizer sampling procedure of the QuSquare benchmark suite~\cite{aguirre2025qusquare} and Eq.~\ref{eq:Y}. For each register size, $l$ non-identity GHZ stabilizers are sampled uniformly at random, each measured $m_i$ times, so shots $=l\times m_i$. Here $l = \lceil 8\ln(4/\delta)/\varepsilon^2\rceil$ with $\varepsilon=\delta=0.10$, giving $l=2952$. Here $m_i=10$ (not the usual $1$), so shots $=29520$ for every row -- this run used a higher \texttt{min\_shots} than Garnet/Emerald/Ibex-Q1. However, only the first measurement outcome for each sampled stabilizer was used to compute the fidelity estimator, resulting in $2952$ measurement outcomes used per GHZ-state size. A register passes when the estimated fidelity satisfies the non-exclusion criterion of Eq.~\eqref{eq:criterion}, $Y\geq 0.5-\varepsilon=0.40$. $^\dagger$Depth/2Q gates are computed on the \textbf{logical} (pre-transpilation, H+CNOT) circuit saved per run -- not the real hardware-native circuit, which was not recorded per-shot by the execution harness (field \texttt{circuit} is hardcoded to \texttt{"Not available in program\_set"} in \texttt{aws\_benchmark.py}). For reference, the actual native-gate (post-transpilation) circuit at $N=16$ was recovered from a separately saved template (not from the per-shot metadata): depth=55, 2Q gates=15 (vs.\ logical depth=10 and 2Q gates=15), fidelity=0.0210. The substantial increase in depth is consistent with the decomposition of CNOT gates into native CZ gates and single-qubit rotations. Hardware figures are the medians of the backend calibration snapshot dated 2026-07-11 (device location: California, USA), taken shortly before the run (2026-07-11).

\normalsize
\vspace{4pt}
\hrule height 0.35pt

\begin{center}
{\normalsize\bfseries Submit Notes}
\end{center}
\vspace{-9pt}
\hrule height 0.35pt
\vspace{3pt}

\noindent\scriptsize\ttfamily
*These experiments were executed using the configuration files defined in \url{https://github.com/NQUIRE-Center/Cross-Platform-Benchmarking/tree/main/GHZ-State_Generation}.\\
The GHZ experiments were executed through Amazon Braket using the Amazon Braket SDK's \texttt{ProgramSet}/\texttt{CircuitBinding} API.\\
For each register size a single run was performed; the GHZ-state preparation circuit (linear H+CNOT ladder) was submitted to the backend, which compiled it server-side to its native gate set.\\
\normalfont

\vspace{2pt}
\end{minipage}}
             
\clearpage

\subsection{Processor \texttt{aqt/Ibex-Q1}}  
\vspace{1cm}

\thispagestyle{empty}
\sffamily\small

\noindent\fbox{%
\begin{minipage}{\dimexpr\textwidth-2\fboxsep-2\fboxrule\relax}
\vspace{1pt}

\begin{center}
{\large\bfseries SPEQC / GHZ Benchmark Report}
\end{center}
\vspace{-4pt}
\hrule height 0.35pt
\vspace{3pt}

\noindent
AWS Braket -- \texttt{qpu/aqt/Ibex-Q1}\\
AQT (Alpine Quantum Technologies)\\
(Trapped-ion QPU, 12 qubits)

\vspace{3pt}
\hrule height 0.35pt
\vspace{3pt}

\noindent
\begin{tabular*}{\linewidth}{@{\extracolsep{\fill}} l l l l @{}}
\textbf{IR Language support:} & OpenQASM 3 & \textbf{Test date:}             & 2026-07-27 \\
\textbf{Test sponsor:}        & AWS& \textbf{Hardware availability:} & 02/26 - \\
\textbf{Tested by:}           & Research team~\cite{nquire2026benchmarkingcode} & \textbf{Software version:} & 1.121.0 \\
\end{tabular*}

\vspace{4pt}

\noindent
\begin{minipage}[t]{0.49\linewidth}
\centering
{\normalsize\bfseries Hardware}\\[-2pt]
\dotfill
\end{minipage}\hfill
\begin{minipage}[t]{0.49\linewidth}
\centering
{\normalsize\bfseries Software}\\[-2pt]
\dotfill
\end{minipage}

\vspace{1pt}

\noindent
\begin{minipage}[t]{0.49\linewidth}
\begin{tabular}{@{}l l@{}}
\textbf{Processor name:}            & AQT Ibex-Q1 \\
\textbf{Backend name:}              & \texttt{aqt/Ibex-Q1} \\
\textbf{Location:}                  & Innsbruck, Austria \\
\textbf{Native gates:}              & PRX, XX, RZ \\
\textbf{2Q gate (median):}          & XX, err 1.57\%, 335 $\mu$s \\
\textbf{1Q gate (median):}          & PRX, err 0.029\%, 45 $\mu$s \\
\textbf{Topology:}                  & Fully connected (all-to-all) \\
\textbf{Total qubits:}              & 12 \\
\textbf{Qubits tested (GHZ size):}  & 7, 10, 11, 12 \\
\textbf{Median T1:}                 & 1.168 s \\
\textbf{Median T2:}                 & 163.2 ms \\
\end{tabular}
\end{minipage}\hfill
\begin{minipage}[t]{0.49\linewidth}
\begin{tabular}{@{}l l@{}}
\textbf{Framework:}        & Amazon Braket SDK \\
\textbf{Runtime:}          & AWS Braket \\
\textbf{Transpiler:}       & Default provider compilation \\
\textbf{Optimization level:} & Not specified  \\
\textbf{Execution mode:}   & On-demand (ProgramSet) \\
\textbf{System state:}     & Multi-user quantum backend \\
\textbf{Repository:}       & \texttt{Cross-Platform-Benchmarking*} \\
\end{tabular}
\end{minipage}

\vspace{4pt}
\hrule height 0.35pt

\begin{center}
{\normalsize\bfseries Results Table}
\end{center}
\vspace{-9pt}
\hrule height 0.35pt
\vspace{3pt}

\noindent\scriptsize
\textbf{A single run was executed per register size (binary search). Depth and 2Q gates are LOGICAL values (pre-transpilation circuit), not the exact native-gate values executed -- see Note.}
\normalsize

\vspace{3pt}

\scriptsize
\renewcommand{\arraystretch}{1.08}
\noindent
\begin{tabular*}{\linewidth}{@{\extracolsep{\fill}} 
L{1.3cm} 
| C{0.8cm} 
C{1.3cm} 
C{1.5cm} 
| C{1.4cm} 
| C{1.1cm} 
C{1.6cm} 
C{1.0cm} 
@{}}
\textcolor{linkblue}{\underline{\bfseries Benchmark}}
& \textcolor{linkblue}{\bfseries Qubits}
& \textcolor{linkblue}{\makecell{\bfseries Depth$^\dagger$}}
& \textcolor{linkblue}{\makecell{\bfseries 2Q gates$^\dagger$}}
& \textcolor{linkblue}{\makecell{\bfseries GHZ\\\bfseries fidelity}}
& \textcolor{linkblue}{\makecell{\bfseries Shots}}
& \textcolor{linkblue}{\makecell{\bfseries Target\\\bfseries threshold}}
& \textcolor{linkblue}{\makecell{\bfseries Pass /\\\bfseries Fail}}
\\
\hline
GHZ & 7 & 5 & 6 & \textbf{0.8763} & 2952 & 0.40 & Pass \\
GHZ & 10 & 6 & 9 & \textbf{0.8369} & 2952 & 0.40 & Pass \\
GHZ & 11 & 6 & 10 & \textbf{0.8165} & 2952 & 0.40 & Pass \\
GHZ & 12 & 6 & 11 & \textbf{0.7114} & 2952 & 0.40 & Pass \\
\hline
\multicolumn{4}{r|}{\textbf{Total circuits executed}} & \multicolumn{4}{c}{4} \\
\multicolumn{4}{r|}{\textbf{Total shots executed}}    & \multicolumn{4}{c}{11\,808} \\
\hline
\multicolumn{8}{c}{\itshape Results shown in ascending register-size order, which matches execution order.} \\
\end{tabular*}

\normalsize
\vspace{4pt}

\noindent
\scriptsize
\textit{Note.}  The reported GHZ fidelity corresponds to the fidelity estimator $Y$, computed from the sample mean $\hat{\mu}$ following the QuSquare benchmark suite~\cite{aguirre2025qusquare} and Eq.~\ref{eq:Y}. The estimate is obtained from a single run per register size. For each size, $l = \lceil 8\ln(4/\delta)/\varepsilon^2\rceil = 2952$ stabilizers of the GHZ state are sampled uniformly at random and each one is measured with a single shot; the 2952 measurement settings are submitted as one parametrized circuit, so each register size accounts for 2952 circuit executions and 2952 shots ($\varepsilon=\delta=0.10$). A register passes when the estimated fidelity satisfies the non-exclusion criterion of Eq.~\eqref{eq:criterion}, $Y\geq 0.5-\varepsilon=0.40$. $^\dagger$Depth/2Q gates are computed on the \textbf{logical} (pre-transpilation, H+CNOT) circuit saved per run -- not the real hardware-native circuit, which was not recorded per-shot by the execution harness (field \texttt{circuit} is hardcoded to \texttt{"Not available in program\_set"} in \texttt{aws\_benchmark.py}). No native (post-transpilation) circuit template was saved for this backend -- unlike Emerald/Cepheus, no native-vs-logical comparison is possible here. Hardware figures are the medians of the backend calibration snapshot dated 2026-07-21 (device location: Innsbruck, Austria), taken shortly before the run (2026-07-27).

\normalsize
\vspace{4pt}
\hrule height 0.35pt

\begin{center}
{\normalsize\bfseries Submit Notes}
\end{center}
\vspace{-9pt}
\hrule height 0.35pt
\vspace{3pt}

\noindent\scriptsize\ttfamily
*These experiments were executed using the configuration files defined in \url{https://github.com/NQUIRE-Center/Cross-Platform-Benchmarking/tree/main/GHZ-State_Generation}.\\
The GHZ experiments were executed through Amazon Braket using the Amazon Braket SDK's \texttt{ProgramSet}/\texttt{CircuitBinding} API.\\
For each register size, a single run was performed. The GHZ-state preparation circuit (linear H+CNOT ladder) was submitted to the backend using the default provider compilation to its native gate set.\\
\normalfont

\vspace{2pt}
\end{minipage}}

\clearpage

\subsection{Processor \texttt{qmio}}
\vspace{1cm}

\thispagestyle{empty}
\sffamily\small

\noindent\fbox{%
\begin{minipage}{\dimexpr\textwidth-2\fboxsep-2\fboxrule\relax}
\vspace{1pt}

\begin{center}
{\large\bfseries SPEQC / GHZ Benchmark Report}
\end{center}
\vspace{-4pt}
\hrule height 0.35pt
\vspace{3pt}

\noindent
CESGA -- \texttt{qmio}\\
Qmio\\
(Superconducting processor, 32 qubits)

\vspace{3pt}
\hrule height 0.35pt
\vspace{3pt}

\noindent
\begin{tabular*}{\linewidth}{@{\extracolsep{\fill}} l l l l @{}}
\textbf{IR Language support:} & OpenQASM 3 & \textbf{Test date:}             & 2026-07-22 / 2026-07-27 \\
\textbf{Test sponsor:}        & CESGA & \textbf{Hardware availability:} & 10/23 - \\
\textbf{Tested by:}           & Research team~\cite{nquire2026benchmarkingcode} & \textbf{Software version:} & 0.21 and 0.5.1 (qmiorun) \\
\end{tabular*}

\vspace{4pt}

\noindent
\begin{minipage}[t]{0.49\linewidth}
\centering
{\normalsize\bfseries Hardware}\\[-2pt]
\dotfill
\end{minipage}\hfill
\begin{minipage}[t]{0.49\linewidth}
\centering
{\normalsize\bfseries Software}\\[-2pt]
\dotfill
\end{minipage}

\vspace{1pt}

\noindent
\begin{minipage}[t]{0.49\linewidth}
\begin{tabular}{@{}l l@{}}
\textbf{Processor name:}            & Qmio \\
\textbf{Backend name:}              & \texttt{qmio} \\
\textbf{Location:}                  & Santiago de Compostela, Spain \\
\textbf{Native gates:}              & ECR, RZ, SX, X \\
\textbf{2Q gate (median):}          & ECR, err 6.411\%, 150 ns \\
\textbf{1Q gate (median):}          & SX, err 0.4335\%, 30 ns \\
\textbf{Topology:}                  & Sparse graph ($\langle$deg$\rangle$=2.00) \\
\textbf{Total qubits:}              & 32 \\
\textbf{Qubits tested (GHZ size):}  & 3, 4, 5, 8, 16 \\
\textbf{Median T1:}                 & 54.1 $\mu$s\textsuperscript{*} \\
\textbf{Median T2:}                 & 77.1 $\mu$s \\
\end{tabular}
\end{minipage}\hfill
\begin{minipage}[t]{0.49\linewidth}
\begin{tabular}{@{}l l@{}}
\textbf{Framework:}        & Qiskit 2.2.3 \\
\textbf{Runtime:}          & qmiotools / QmioBackend \\
\textbf{Transpiler:}       & Qiskit transpiler \\
\textbf{Optimization level:} & 3 \\
\textbf{Execution mode:}   & vqc (RS benchmark) \\
\textbf{System state:}     & Multi-user quantum backend \\
\textbf{Repository:}       & \texttt{Cross-Platform-Benchmarking*} \\
\end{tabular}
\end{minipage}

\vspace{4pt}
\hrule height 0.35pt

\begin{center}
{\normalsize\bfseries Results Table}
\end{center}
\vspace{-9pt}
\hrule height 0.35pt
\vspace{3pt}

\noindent\scriptsize
\textbf{A single run was executed per register size (binary search over 2--32 qubits). Depth and 2Q gate counts are exact values, read directly from the executed OpenQASM~3 circuit (not reconstructed).}
\normalsize

\vspace{3pt}

\scriptsize
\renewcommand{\arraystretch}{1.08}
\noindent
\begin{tabular*}{\linewidth}{@{\extracolsep{\fill}} 
L{1.3cm} 
| C{0.8cm} 
C{1.0cm} 
C{1.3cm} 
| C{1.4cm} 
| C{1.1cm} 
C{1.6cm} 
C{1.0cm} 
@{}}
\textcolor{linkblue}{\underline{\bfseries Benchmark}}
& \textcolor{linkblue}{\bfseries Qubits}
& \textcolor{linkblue}{\makecell{\bfseries Depth}}
& \textcolor{linkblue}{\makecell{\bfseries 2Q gates}}
& \textcolor{linkblue}{\makecell{\bfseries GHZ\\\bfseries fidelity}}
& \textcolor{linkblue}{\makecell{\bfseries Shots}}
& \textcolor{linkblue}{\makecell{\bfseries Target\\\bfseries threshold}}
& \textcolor{linkblue}{\makecell{\bfseries Pass /\\\bfseries Fail}}
\\
\hline
GHZ & 3  & 15 & 2  & \textbf{0.5127}  & 2952 & 0.40 & Pass \\
GHZ & 4  & 15 & 3  & \textbf{0.3769}  & 2952 & 0.40 & \textbf{Fail} \\
GHZ & 5  & 12 & 4  & \textbf{0.1402}  & 2952 & 0.40 & \textbf{Fail} \\
GHZ & 8  & 15 & 7  & \textbf{0.0161}  & 2952 & 0.40 & \textbf{Fail} \\
GHZ & 16 & 25 & 15 & \textbf{$-0.0108$} & 2952 & 0.40 & \textbf{Fail} \\
\hline
\multicolumn{4}{r|}{\textbf{Total circuits executed}} & \multicolumn{4}{c}{5} \\
\multicolumn{4}{r|}{\textbf{Total shots executed}}    & \multicolumn{4}{c}{14\,760} \\
\hline
\multicolumn{8}{c}{\itshape Sorted by register size. Execution order (binary search): 16, 8, 5, 4, 3.} \\
\end{tabular*}

\normalsize
\vspace{4pt}

\noindent
\scriptsize
\textit{Note.} 
The reported GHZ fidelity corresponds to the fidelity estimator $Y$, computed from the sample mean $\hat{\mu}$ following the QuSquare benchmark suite~\cite{aguirre2025qusquare} and Eq.~\ref{eq:Y}. The estimate is obtained from a single run per register size. For each size, $l = \lceil 8\ln(4/\delta)/\varepsilon^2\rceil = 2952$ stabilizers of the GHZ state are sampled uniformly at random and each one is measured with a single shot; the 2952 measurement settings are submitted as one parametrized circuit, so each register size accounts for 2952 circuit executions and 2952 shots ($\varepsilon=\delta=0.10$). A register passes when the estimated fidelity satisfies the non-exclusion criterion of Eq.~\eqref{eq:criterion}, $Y\geq 0.5-\varepsilon=0.40$.  Circuit depth and 2Q gate counts are exact: read directly from the transpiled OpenQASM~3 circuit (\texttt{GHZ\_\{n\}q.qasm}) that was executed. Registers were run on 2026-07-22 (3q, 5q, 8q, 16q) and 2026-07-27 (4q). Hardware figures (native gates, ECR/SX/readout medians, T2) were identical across all four saved calibration snapshots. \textsuperscript{*}T1 was identical (54.1\,$\mu$s) in the three 2026-07-22 snapshots; the 2026-07-27 snapshot (used for the 4-qubit run) measured 50.5\,$\mu$s instead. Median T2 is taken over the 21/32 qubits reporting valid values (11 qubits stored placeholder $\sim$1e-15\,s entries, excluded).

\normalsize
\vspace{4pt}
\hrule height 0.35pt

\begin{center}
{\normalsize\bfseries Transpiler notes}
\end{center}
\vspace{-9pt}
\hrule height 0.35pt
\vspace{3pt}

\noindent\scriptsize\ttfamily
The Qiskit transpiler is available at \url{https://github.com/Qiskit/qiskit}
\normalfont

\vspace{4pt}
\hrule height 0.35pt

\begin{center}
{\normalsize\bfseries Submit Notes}
\end{center}
\vspace{-9pt}
\hrule height 0.35pt
\vspace{3pt}

\noindent\scriptsize\ttfamily
*These experiments were executed using the configuration files defined in \url{https://github.com/NQUIRE-Center/Cross-Platform-Benchmarking/tree/main/GHZ-State_Generation}.\\
The GHZ experiments were executed on the real Qmio backend (\texttt{QmioBackend}) via SLURM on the qmio/hpc partition.\\
For each register size a single run was performed; the GHZ-state preparation circuit was transpiled (opt.\ level 3) to the native gate set (ECR, RZ, SX, X) and connectivity of the backend, then executed under the RS benchmark in vqc mode.\\
The reported fidelity is the random-stabilizer-sampling estimator value obtained for each register size.
\normalfont

\vspace{2pt}
\end{minipage}}

\clearpage

\section{Rigetti Cepheus-1 with manually selected qubits}\label{ceph}

As discussed in the main text, the default provider compiler for \texttt{rigetti\_cepheus} selected a faulty physical qubit during our benchmark experiments. To investigate the influence of this qubit assignment on the observed performance, we repeated the benchmark using manually selected physical qubits based on calibration data. This experiment complements the default compilation results reported in the main text by explicitly incorporating single- and two-qubit gate fidelities into the qubit selection process.

The physical qubits were selected using a graph-based optimization procedure that identifies connected chains of physical qubits using the calibration data provided by the device. The processor is represented by an undirected graph $G=(V,E)$, where each node $i\in V$ corresponds to a physical qubit with a valid single-qubit gate fidelity $f_i$, and each edge $(i,j)\in E$ represents an available physical connection with a valid CZ-gate fidelity $f_{ij}$. Qubits without valid single-qubit calibration data are excluded, as are connections for which the corresponding CZ-gate fidelity is unavailable or invalid.

The calibration information is converted into a cost function using negative logarithms of the gate fidelities. Specifically, the weight assigned to an edge is
\begin{equation}
    w_{ij}
    =
    -\log f_{ij}
    -\log f_i
    -\log f_j.
    \label{eq:edge_cost}
\end{equation}
This expression provides an additive proxy for the multiplicative effect of the fidelities associated with a two-qubit operation and its endpoint qubits. For a chain $P=(v_1,\ldots,v_N)$, the optimization procedure minimizes
\begin{equation}
\begin{aligned}
    C(P) ={}&
    -\alpha \log f_{v_1}\\
    &+\sum_{k=1}^{N-1} w_{v_k v_{k+1}}\\
    &-\alpha\sum_{k=2}^{N}\log f_{v_k},
\end{aligned}
\label{eq:chain_cost}
\end{equation}
where $\alpha=0.25$ is the weight assigned to the additional single-qubit contribution in the implementation. The chain is required to contain exactly $N$ distinct physical qubits, and only edges present in the calibration graph are considered. The search enumerates candidate paths and uses branch-and-bound pruning to discard partial paths whose accumulated cost is no smaller than the best complete solution found so far.

The calibration-aware selection procedure was evaluated on Rigetti Cepheus-1 using GHZ circuits with different numbers of physical qubits. Table~\ref{tab:manual_selection_results} summarizes the measured GHZ-state fidelities obtained with the new selection strategy.

\begin{table}[ht]
    \centering
    \caption{GHZ-state fidelities obtained with the calibration-aware qubit selection.}
    \label{tab:manual_selection_results}
    \begin{tabular}{lc}
        \hline
        GHZ size $N$ & GHZ fidelity $Y$\\
        \hline
        2  & 0.86\\
        3  & 0.42\\
        4  & 0.33\\
        8  & 0.25\\
        16 & 0.04\\
        \hline
    \end{tabular}
\end{table}

For the 16-qubit chain, the optimization reduced the objective value from 1.88 to 0.15, corresponding to a reduction of approximately 92\%. Nevertheless, the measured GHZ-state fidelity was only 0.04. Despite the more favorable calibration-based costs of the selected qubits and connections, the results did not show the expected improvement. The estimated GHZ-state fidelities for the other tested sizes were 0.25, 0.33, 0.86, and 0.42 for $N=8$, $4$, $2$, and $3$, respectively. Under the benchmark criterion $Y+\epsilon\geq 1/2$ with $\epsilon=0.1$, the three-qubit circuit satisfies the criterion, whereas the four-qubit circuit does not. The largest GHZ state satisfying the benchmark criterion was therefore $N=3$, compared with $N^*=4$ obtained using the default compilation workflow.

Although the default provider compiler selected a physical qubit with poor calibration fidelity during our experiments, replacing the default qubit assignment with a chain chosen to minimize a calibration-based cost did not improve the largest GHZ-state satisfying the benchmark criterion. Despite a reduction of approximately one order of magnitude in the estimated cost for the 16-qubit circuit, the estimated GHZ-state fidelity remained low. This suggests that the physical qubit selection performed by the default compiler is not the only factor limiting the performance of Rigetti Cepheus-1. Identifying the underlying causes requires a more thorough investigation of hardware-related and other potential factors. Such an investigation is beyond the scope of this work.

\section{Assessment of temporal variability in repeated executions}
\label{temp_var}

Ideally, quantum benchmarking experiments should be repeated at different points in time, and their results averaged to account for temporal variations in processor calibration and operating conditions. However, the cost of accessing cloud-based quantum processors imposes significant constraints on the number of experimental executions, alongside practical limitations related to execution time and hardware availability. The experimental budget therefore required a trade-off between repeating benchmarks executions on the same processors and extending the evaluation to a broader set of quantum processors. We prioritized the latter to increase hardware coverage.

Nevertheless, to assess the temporal stability of the benchmark, we performed additional executions on a subset of processors at different points in time. These repeated executions allowed us to examine variations in the estimated GHZ fidelities and maximum GHZ-state sizes identified by the benchmark. For this temporal-variability analysis, we used $\epsilon=\delta=0.01$, imposing a stricter statistical accuracy target and confidence level than in the main experiments. This reduces the contribution of statistical estimation error to differences between paired executions, making the comparison more informative about variability beyond finite-sampling effects.

\begin{figure}[htbp]
    \centering
    \includegraphics[width=0.8\textwidth]{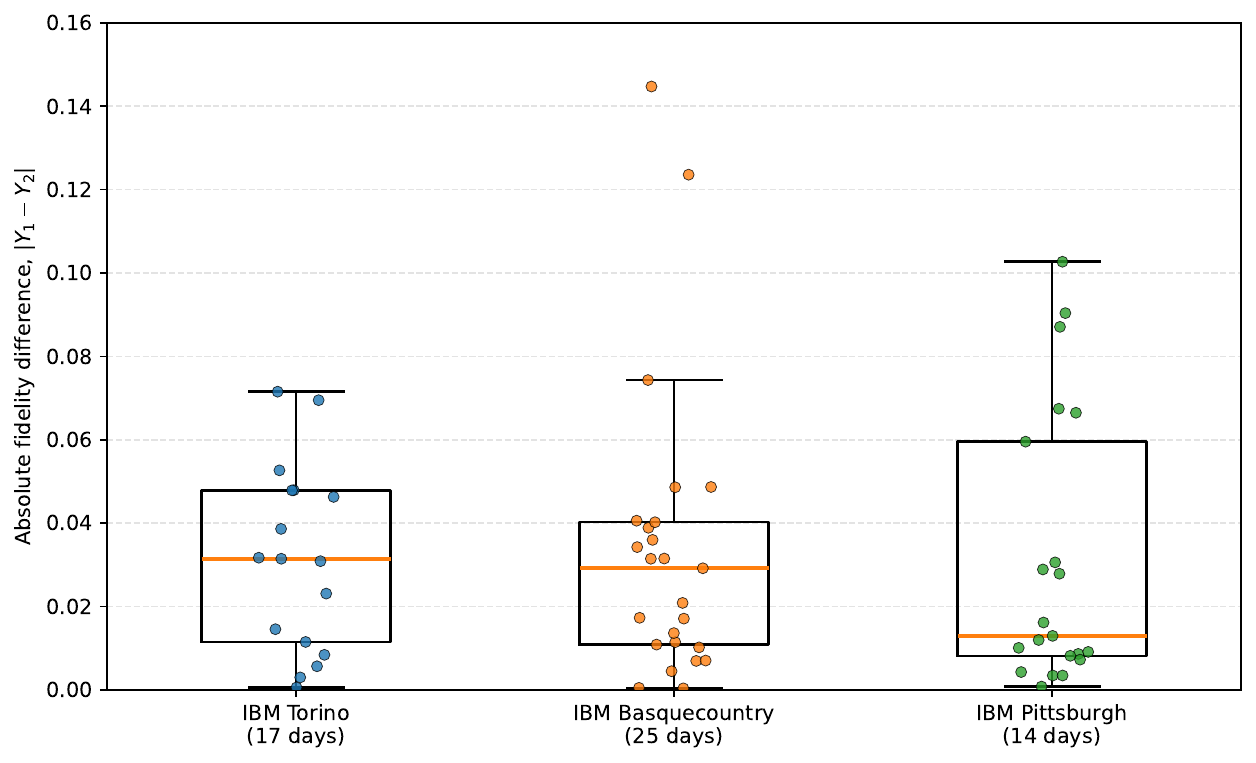}
    \caption{Absolute differences in GHZ fidelity estimates, $|Y_1-Y_2|$, between executions performed at different times on three IBM Quantum processors. Each point corresponds to a GHZ-state size tested in both executions, and the box plots summarize the distribution of absolute differences for each processor. The executions were separated by 17 days for \texttt{ibm\_torino}, 25 days for \texttt{ibm\_basquecountry}, and 14 days for \texttt{ibm\_pittsburgh}. The \texttt{ibm\_pittsburgh} observation (0.4443) was excluded because it compared different GHZ-state sizes across the two executions.}
    \label{fig:fidelity_variations}
\end{figure}

For each GHZ-state size tested in both executions, we calculated the absolute difference between the corresponding fidelity estimates, ($|Y_1-Y_2|$). Figure \ref{fig:fidelity_variations} presents the resulting distributions. Although many differences were below 0.05, some substantially larger discrepancies were observed, particularly for ibm\_basquecountry and ibm\_pittsburgh. These results indicate that fidelity estimates can exhibit appreciable run-to-run variability and that its magnitude depends on the processor and experimental instance.

In contrast, the maximum GHZ-state size achieved was considerably more consistent across the paired executions: only one comparison resulted in a different maximum size, and the difference was one qubit. This suggests that, under the conditions tested, the maximum GHZ-state size identified by the benchmark is less sensitive to temporal variability than the individual fidelity estimates.

The observed differences should be interpreted in the context of the time-dependent operating conditions of cloud-based QPUs. Calibration parameters, qubit mappings, and other operational settings may vary between executions, potentially causing performance fluctuations even when the underlying hardware remains unchanged. Such variations can arise, for example, when a device is measured at different stages of its calibration cycle. In the comparisons considered, the maximum GHZ-state size identified by the benchmark remained largely stable, although individual fidelity estimates exhibited appreciable variations.

A distinct source of temporal change is the evolution of the platform itself. Providers may deploy hardware upgrades, control-system improvements, or software and compilation updates that systematically change benchmark outcomes. These changes should not be conflated with short-term operational variability: they may represent genuine improvements, regressions, or trade-offs in device performance. Although the present experiments do not allow us to identify or quantify specific provider-side changes, they highlight the importance of periodically rerunning a consistent, reproducible benchmark. Making the benchmark methodology publicly available allows independent users to monitor performance over time, investigate observed changes, and assess whether platform updates translate into measurable improvements in application-relevant performance.

\end{document}